\documentclass{aastex631}

\usepackage{amsmath}
\usepackage{tasks}
\usepackage{booktabs}
\usepackage{rotating}
\usepackage{makecell}
\usepackage{xspace}
\usepackage{enumitem}
\usepackage{textcomp}
\usepackage{stackengine}
\usepackage{bm}
\usepackage[caption=false]{subfig}
\setlist[itemize]{noitemsep}
\usepackage{array}

\newcommand{\PUCChile}{Institute of Astrophysics, Pontificia Universidad Cat\'olica de Chile, Santiago, Chile}

\newcommand{\OxfordAstro}{Astrophysics Sub-department, University of Oxford, Oxford, UK}

\newcommand{\USF}{Department of Physics and Astronomy, University of San Francisco, San Francisco, CA 94117, USA}

\newcommand{\LBNL}{Physics Division, Lawrence Berkeley National Laboratory, Berkeley, CA 94720, USA}

\newcommand{\UCLA}{Department of Physics and Astronomy, University of California, Los Angeles, Los Angeles, CA 90095, USA}

\newcommand{\SCIPP}{Santa Cruz Institute for Particle Physics, University of California, Santa Cruz, Santa Cruz, CA 95064, USA}

\newcommand{\UCBPhysics}{Department of Physics, University of California, Berkeley, Berkeley, CA 94720, USA}

\newcommand{\UChicago}{Department of Astronomy, University of Chicago, Chicago, IL 60637, USA}

\newcommand{\UCMadrid}{Department of Physics, Complutense University of Madrid, 28040 Madrid, Spain}

\newcommand{\Swinburne}{Centre for Astrophysics and Supercomputing, Swinburne University of Technology, Hawthorn, Victoria 3122, Australia}

\newcommand{\ASTROThreeD}{ARC Centre of Excellence for All Sky Astrophysics in 3 Dimensions (ASTRO 3D), Australia}

\newcommand{\CfA}{Center for Astrophysics, Harvard \& Smithsonian, Cambridge, MA 02138, USA}

\newcommand{\ESOChile}{European Southern Observatory, Alonso de C\'ordova 3107, Vitacura, Santiago, Chile}

\newcommand{\SLAC}{Fundamental Physics Directorate, SLAC National Accelerator Laboratory, Menlo Park, CA 94025, USA}

\newcommand{\GeminiObs}{Gemini Observatory, NSF's NOIRLab, La Serena, Chile}

\newcommand{\BCCP}{Berkeley Center for Cosmological Physics, University of California, Berkeley, Berkeley, CA 94720, USA}

\newcommand{\UCBAstro}{Department of Astronomy, University of California, Berkeley, Berkeley, CA 94720, USA}

\newcommand{\UCDavis}{Department of Physics and Astronomy, University of California, Davis, Davis, CA 95616, USA}

\newcommand{\UHM}{Department of Physics and Astronomy, University of Hawai\textquoteright i at M\=anoa, Honolulu, HI 96822, USA}

\newcommand{\HarvardPhysics}{Department of Physics, Harvard University, Cambridge, MA 02138, USA}

\newcommand{\IfAUH}{Institute for Astronomy, University of Hawai\textquoteright i, Honolulu, HI 96822, USA}

\newcommand{\FSU}{Department of Physics, Florida State University, Tallahassee, FL 32306, USA}

\newcommand{\KavliTokyo}{Kavli Institute for the Physics and Mathematics of the Universe (WPI), University of Tokyo, Kashiwa 277-8583, Japan}

\shorttitle{Cosmological Constraints from the Carousel Lens with \gigal}
\shortauthors{}

\graphicspath{{./}{figures/}}

\newcommand{\gigal}{\texttt{GIGA-Lens}\xspace}

\newcommand\angdot[1][\circ]{%
  \stackengine{0pt}{.}{${}^{\mathrm{#1}}$}{O}{l}{F}{F}{L}}
\newcommand\angdotcustom[1][\circ]{%
  \stackengine{0pt}{.}{${\mathrm{#1}}$}{O}{l}{F}{F}{L}}
  
\newcolumntype{L}[1]{>{\arraybackslash}p{#1}}

\begin{document}

\title{The Carousel Lens II: Cosmological Constraints with \texttt{GIGA-Lens}}



\author[0009-0009-0407-2419]{Felipe~Urcelay}
\affiliation{\PUCChile}
\affiliation{\OxfordAstro}

\author[0000-0001-8156-0330]{Xiaosheng~Huang}
\affiliation{\USF}
\affiliation{\LBNL}

\author[0000-0003-1889-0227]{William~Sheu}
\affiliation{\UCLA}

\author[0000-0003-4083-1530]{Jackson~H.~O'Donnell}
\affiliation{\SCIPP}

\author[0000-0001-6089-0365]{Tesla~Jeltema}
\affiliation{\SCIPP}

\author[0000-0003-1042-1995]{Demetrius~Y.~Williams}
\affiliation{\SCIPP}

\author[0009-0006-4623-3629]{Sean~Xu}
\affiliation{\UCBPhysics}


\author[0000-0002-2350-4610]{Shrihan~Agarwal}
\affiliation{\UChicago}

\author{Greg~Aldering}
\affiliation{\LBNL}

\author[0009-0006-0284-3863]{David~\'{A}lvarez-Garc\'{i}a}
\affiliation{\LBNL}
\affiliation{\UCMadrid}

\author[0009-0003-5592-3515]{Harsh~Ambardekar}
\affiliation{\UCBPhysics}

\author[0000-0002-2784-564X]{Tania~M.~Barone}
\affiliation{\Swinburne}
\affiliation{\ASTROThreeD}
\affiliation{\CfA}

\author[0000-0002-1620-0897]{Fuyan~Bian}
\affiliation{\ESOChile}

\author[0000-0002-9836-603X]{Adam~S.~Bolton}
\affiliation{\SLAC}

\author[0000-0001-7101-9831]{Aleksandar~Cikota}
\affiliation{\GeminiObs}

\author[0000-0001-5704-1127]{Gerrit~S. Farren}
\affiliation{\LBNL}
\affiliation{\BCCP}

\author[0000-0002-3254-9044]{Karl~Glazebrook}
\affiliation{\Swinburne}
\affiliation{\ASTROThreeD}

\author[0000-0001-9664-0560]{Taylor~Hoyt}
\affiliation{\LBNL}
\affiliation{\UCBPhysics}

\author[0009-0002-6935-3763]{Aniket~Jain}
\affiliation{\UCBAstro}

\author[0000-0001-5860-3419]{Tucker~Jones}
\affiliation{\UCDavis}

\author[0000-0003-1362-9302]{Glenn~G.~Kacprzak}
\affiliation{\Swinburne}
\affiliation{\ASTROThreeD}

\author[0009-0006-5989-4899]{Emerald~Lin}
\affiliation{\UCBPhysics}

\author[0000-0002-4436-4661]{Saul~Perlmutter}
\affiliation{\LBNL}
\affiliation{\UCBPhysics}


\author[0000-0001-5402-4647]{David Rubin}
\affiliation{\UHM} 
\affiliation{\LBNL}

\author[0000-0002-5042-5088]{David~J.~Schlegel}
\affiliation{\LBNL}

\author[0000-0002-1804-3960]{Ethan~Silver}
\affiliation{\HarvardPhysics}

\author[0000-0002-0385-0014]{Christopher~J.~Storfer}
\affiliation{\LBNL}
\affiliation{\IfAUH}

\author[0000-0001-7266-930X]{Nao~Suzuki}
\affiliation{\LBNL}
\affiliation{\FSU}
\affiliation{\KavliTokyo}

\author[0009-0004-8106-9452]{Jannik~Truong}\affiliation{\UCBPhysics}

\author[0009-0005-4355-0293]{M\'{o}nica~\'{U}beda}
\affiliation{\LBNL}
\affiliation{\UCMadrid}

\author[0000-0002-2645-679X]{Keerthi~Vasan~G.~C.}
\affiliation{\UCDavis}

\begin{abstract}

The nature of dark matter and dark energy are among the central questions in cosmology. Strong gravitational lenses with multiple source planes provide a geometric probe of cosmology: the ratio of deflection angles at different redshifts depends only on angular-diameter distances, constraining the matter density $\Omega_m$ and the dark energy equation of state $w$. However, constraints from this technique have historically lagged behind those from the CMB, SNe~Ia, and BAO. In this work, we present new cosmological constraints from the Carousel Lens, a cluster-scale lens with more than 40 extended images from 11 spectroscopically confirmed sources. Its relaxed core and rich set of extended images behind the main halo make it particularly suitable for cosmological inference. Using the \texttt{GIGA-Lens} pipeline, we construct a pixel-level lens model including six HST-detected sources and four mass components. From this model, we obtain $w$CDM constraints of $\Omega_m = 0.34^{+0.16}_{-0.13}$ and $w = -1.31^{+0.35}_{-0.32}$ from the Carousel Lens alone, accounting for both statistical and systematic uncertainties. We further project that including four additional known higher-redshift sources, assuming similar fractional uncertainties, could improve the constraining power by $\sim80\%$, bringing the precision close to that of the CMB and SNe Ia. For an evolving dark energy model ($w_0w_a$CDM), the Carousel Lens alone yields constraints comparable to the CMB, providing an independent and complementary probe alongside SN~Ia and BAO. While currently systematic uncertainties dominate, which we quantify through simulations, our results demonstrate that relaxed multi-source-plane cluster lenses can deliver competitive cosmological constraints. Further improvements are expected from reductions in systematics and from incorporating higher-redshift sources (known and new) with high-resolution imaging.

\end{abstract}

\keywords{Strong gravitational lensing --- Observational cosmology --- Dark matter density --- Dark energy --- Galaxy clusters}

\section{Introduction} \label{sec:introduction}
The $\Lambda$ cold dark matter ($\Lambda$CDM) cosmological model has proven highly successful at describing the large-scale properties of the Universe, which until recently was supported by a broad range of observations including the cosmic microwave background (CMB), baryon acoustic oscillations (BAO), type Ia supernovae (SNe~Ia), and the large-scale structure \citep[e.g.,][]{planck2018, Aubourg2015, Alam2017, Brout2022, Abbott2022}. In this framework, roughly 30\% of the cosmic energy budget is in matter, predominantly cold dark matter, while the remaining $\sim$70\% is in dark energy, modeled as a cosmological constant, that drives the accelerated expansion of the Universe. Yet, despite this concordance, growing tensions have emerged between independent measurements of key parameters, most notably the Hubble constant \citep{planck2018, Brout2022}, and uncertainties remain on the properties of dark energy itself \citep{Zhao2017, Rubin2025, Abdul2025}. These discrepancies motivate the exploration of complementary and independent cosmological probes that are subject to different systematics.  

Strong gravitational lensing provides one such probe. Because the positions, multiplicities, and magnifications of lensed images depend on angular-diameter distance ratios between the observer, lens, and background sources, strong lensing directly encodes information on cosmological parameters. Time delays in variable sources yield constraints on $H_0$ \citep{Refsdal1964b, Kelly2023, Pascale2024, tdcosmo2025}, while multi-source plane systems can, in principle, constrain both the matter density $\Omega_m$ and the dark energy equation of state $w$ \citep{Collett2014, Caminha2022, Bolamperti2024, Sahu2025, Bowden2025}. 

In the context of multi-source-plane cosmography, there are two lensing regimes, each with distinct trade-offs. Galaxy-scale double-source-plane lenses benefit from well-constrained deflectors, but their individual cosmological constraining power is limited \citep[e.g.,][]{Collett2014, Sahu2025, Bowden2025} and must be compensated by combining large samples \citep{Sharma2023}. In addition, the close alignment of the lens galaxy and the two source planes implies that the nearer source can itself act as an additional deflector, introducing multi-plane lensing effects that adds to model complexity. Cluster-scale lenses, on the other hand, provide a much larger number of constraints from multiple image families spanning a wide range of source redshifts, but often require much more complex mass models, typically with a large number of components, and are therefore more susceptible to modeling systematics.

Relatively relaxed group- and cluster-scale lensing systems offer an intermediate regime between galaxy-scale lenses and dynamically complex massive clusters. They often feature more lensed source planes, larger Einstein radii than galaxies, and simpler total mass distributions than most clusters, making them promising candidates for precision cosmography. An example of such a system is the group-scale lens presented by \citet{Bolamperti2024}, with an Einstein radius of $\theta_E \sim 7''$. 

In this work, we present new strong lensing constraints from the Carousel Lens, a cluster lensing system that, to date, is known to generate multiple images of 12 background galaxies, with redshifts ranging from 0.9 to 4.1 \citep{odonnell2026}. Although the cluster is massive ($M_{200c}\sim 10^{15} \, M_\odot$ based on kinematic measurements by \citet{odonnell2026} and consistent with X-ray and Sunyaev–Zel’dovich measurements), \citet{sheu2024a} showed that a comparatively simple mass model---consisting of a single elliptical cluster-scale halo plus one galaxy-scale subhalo---can accurately reproduce the multiple-image configurations observed in the cluster core. This combination of a massive lens, with $\theta_E\sim 13''$ at $z_{source} = 1.4$, a rich set of multiply imaged sources, and a relatively simple inner ($< 30''$) mass distribution makes the Carousel Lens a uniquely powerful system for cosmological inference, offering enhanced constraining power compared to individual galaxy-scale lenses and more complex cluster-scale systems.

In this article, we present a new strong-lensing analysis of the Carousel Lens based on extended surface-brightness modeling at the pixel level, and derive cosmological constraints on $\Omega_m$ and $w$ from its multi-source-plane configuration, including a detailed assessment of systematic uncertainties. Finally, we forecast the constraining power achievable with high-resolution imaging of newly identified higher-redshift sources and discuss the potential improvements to our analysis.

The paper is organized as follows. We describe the theoretical background in Section~\ref{sec:cosmo_theo}, data in Section~\ref{sec:data}, and the lens model in Section~\ref{sec:lens_model}, and proceed to present the cosmological inference in Section~\ref{sec:cosmo} and discuss comparisons and improvements in Section~\ref{sec:discussion}, and conclude in Section~\ref{sec:conclusion}. 

\section{Cosmography with multi-source-plane strong lenses} \label{sec:cosmo_theo}
In gravitational lensing, the mapping between the source plane and the image plane is described by the lens equation \citep{Schneider1992}:
\begin{equation} \label{eq:lens_equation}
    \bm{\beta} = \bm{\theta} - \bm{\alpha}(\bm{\theta})
\end{equation}
where $\bm{\beta}$ is the angular position of the source, $\bm{\theta}$ is the observed image position, and $\bm{\alpha}(\bm{\theta})$ is the reduced deflection angle produced by the lens at position $\bm{\theta}$. This deflection angle is given by:
\begin{equation}
    \bm{\alpha}(\bm{\theta}) =
    \frac{1}{\pi} \int
    \kappa(\bm{\theta}')
    \frac{\bm{\theta} - \bm{\theta}'}{|\bm{\theta} - \bm{\theta}'|^2}
     d^2\bm{\theta}'
\end{equation}
Here, the convergence $\kappa(\bm{\theta}) = \Sigma(D_l \, \bm{\theta}) / \Sigma_{\text{cr}}$ is the surface mass density $\Sigma$ scaled by the critical surface density $\Sigma_{\text{cr}}$, defined as:
\begin{equation}
    \Sigma_{\text{cr}}(z_l, z_s) \equiv \frac{c^2}{4 \pi G} \frac{D_s(z_s)}{D_l(z_l) \, D_{ls}(z_l, z_s)}
\end{equation}
where $D_l$ and $D_s$ are the angular diameter distances from the observer to the lens and source, respectively, and $D_{ls}$ is the distance from the lens to the source, with $z_l$ and $z_s$ denoting the redshifts of the lens and source.

A lens deflects all light rays by a physical deflection angle $\hat{\bm{\alpha}}(\bm{\theta}) = (D_s / D_{ls}) \, \bm{\alpha}(\bm{\theta})$. Therefore, for multiple sources at different redshifts, the reduced deflection angle (henceforth, simply, the deflection angle) for a source at redshift $z_i$ can be obtained by scaling the deflection measured at a reference redshift $z_{\text{ref}}$ by a deflection ratio $\eta_i$, defined as:
\begin{equation} \label{eq:distance_ratio}
    \eta_i(z_l, z_i, z_{\text{ref}}) \equiv 
    \frac{\bm{\alpha}_i}{\bm{\alpha}_{\text{ref}}} = 
    \frac{D_{ls}(z_l, z_{\text{ref}}) \, D_s(z_i)}{D_s(z_{\text{ref}}) \, D_{ls}(z_l, z_i)}
\end{equation}
From a modeling perspective, $\bm{\alpha}_{\text{ref}}$ can be regarded as depending solely on the lens convergence, while $\eta_i$ encodes the dependence on cosmology. To illustrate how $\eta_i$ varies with cosmological parameters, we consider the expression for the angular diameter distance in a flat universe:
\begin{equation}
    D(z_1, z_2) = \frac{c}{H_0(1+z_2)} \int_{z_1}^{z_2} \frac{dz}{\sqrt{\Omega_m(1+z)^3 + \Omega_r(1+z)^4 + (1 - \Omega_m + \Omega_r)(1+z)^{3(1+w)}}}
\end{equation}
where $\Omega_r$ is the density parameter for relativistic particles. In equation~\eqref{eq:distance_ratio}, the Hubble constant $H_0$ cancels out, making $\eta_i$ primarily sensitive to the parameters $\Omega_m$ and $w$. These two parameters are fully degenerate in this context: there are infinitely many combinations of $\Omega_m$ and $w$ that yield the same deflection ratio. This degeneracy defines a curve in the $\Omega_m$–$w$ parameter space, which depends on the specific redshift configuration of the lensing system. With multiple source planes, this degeneracy region around this curve can be significantly reduced.

\section{Data} \label{sec:data}
\begin{figure}[htbp]
    \centering
    \includegraphics[width=0.8\linewidth]{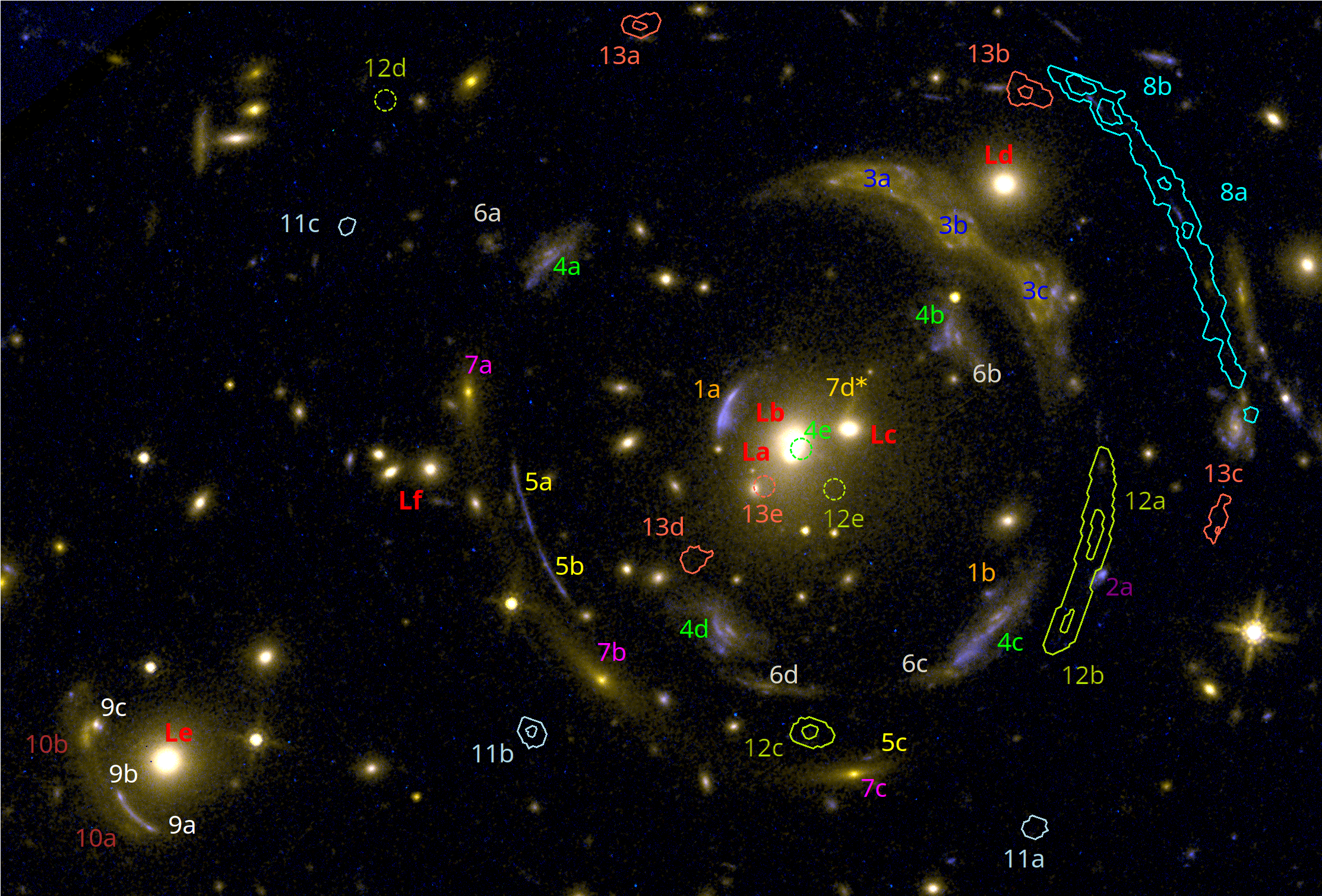}
    \caption{All currently known lensed sources in the Carousel Lens system. 
    The images of the same source galaxy are labeled with the same number and color, with the letters indicating the image multiplicity of each source family. 7d$^*$ is the suspected radial image, but it has not been spectroscopically confirmed.
    Sources with Ly$\alpha$ emission detected in the MUSE data are shown with contours. The most prominent cluster members are labeled in red. This figure is reproduced from Paper~I.}
    \label{fig:all-sources-jack}
\end{figure}

\begin{table}[htbp]
    \centering
    \caption{Sources redshifts and image multiplicities.}
    \begin{tabular}{ccccc}
        \hline
        \hline
        Source & redshift & multiplicity & modeled & reference \\
        \hline
        1 & 0.962 & 2 & y & 1 \\
        2 & 0.962 & 1 & n & 1\\
        3 & 1.166 & 3 & y & 1\\
        4 & 1.432 & 5$^a$ & y & 1\\
        5 & 1.432 & 3 & y & 1\\
        6 & 1.656 & 4 & y & 2 \\
        7 & 1.627 & 4$^b$ & y & 2\\
        8 & 3.549 & 3 & n & 2 \\
        9 & 1.507 & 3 & y$^c$ & 2\\
        10 & - & 3 & n & - \\
        11 & 4.09 & 3 & n & 2 \\
        12 & 3.086 & 5 & n & 2 \\
        13 & 3.086 & 5 & n & 2 \\
        \hline
        \multicolumn{5}{L{7.5cm}}{\footnotesize{The table lists the source family ID, spectroscopic redshift, number of images detected in the current data, whether the source is included in the lens model (y/n), and the reference for the redshift measurement.}} \\
        \multicolumn{5}{l}{\footnotesize{1: \citet{sheu2024a}}, Paper~0} \\
        \multicolumn{5}{l}{\footnotesize{2: \citet{odonnell2026}}, Paper~I} \\
        \multicolumn{5}{L{7.5cm}}{\footnotesize{$^a$A fifth inner image is identified in the MUSE data but not visible in HST.}} \\
        \multicolumn{5}{L{7.5cm}}{\footnotesize{$^b$Three of the four images, but not yet the radial arc, identified in HST are spectroscopically confirmed.}} \\
        \multicolumn{5}{L{7.5cm}}{\footnotesize{$^c$Only with the image positions.}} \\
    \end{tabular}
    \label{tab:sources-z}
\end{table}

DESI~J090.9854-35.9683\footnote{The naming convention is RA and Dec in decimal format.} (RA: $6^{\rm h} 3^{\rm m} 56 \angdot[{\rm s}] 50$, DEC.: $-35^{\circ} 58' 5\angdotcustom[{''}] 88$), hereafter referred to as the Carousel Lens, is a galaxy cluster at redshift $z_{\rm l}=0.49$. It was first identified as a strong-lensing candidate by \citet{jacobs2019} using a convolutional neural network applied to Dark Energy Survey (DES) Year~3 imaging data, and subsequently independently discovered by \citet{huang2021} using DESI Legacy Imaging Surveys Data Release~8 and was assigned a Grade-A strong-lens classification.

Follow-up observations with the Hubble Space Telescope (HST) Wide Field Camera~3 (WFC3) in the F200LP and F140W bands (600~s each; Program~ID~16773, PI: K.~Glazebrook), together with shallow VLT/MUSE integral-field spectroscopy (45~min; Program~ID~0111.B-0400(H), PI: A.~Cikota), were presented by \citet{sheu2024a} (Carousel Lens Paper~0). These data enabled a spectroscopic confirmation of the cluster redshift and of four multiply imaged background sources (sources~1, 3, 4, and 5; source~2 is single-imaged), as well as the construction of an initial strong-lensing model. Two additional candidate systems (sources~6 and~7) were identified in the HST imaging based on their colors, but their redshifts remained unconstrained at that stage.

Subsequently, deeper Gemini/GMOS spectroscopy (Program~ID~S-2023B-Q-103, PI: T.~Jeltema) and VLT/MUSE observations (2.76~h; Program~ID~0114.A-2018(A), PI: T.~Barone) were analyzed by \citet{odonnell2026} (Carousel Lens Paper~I). From a kinematic analysis of 49 confirmed cluster members, they estimated a cluster halo mass of $M_{200c}\sim 1.2 \times 10^{15}\,M_\odot$. These data also provided spectroscopic redshifts for sources~6 and~7 (both at $z \simeq 1.6$) and led to the discovery of six additional multiply imaged systems, including four Ly$\alpha$ emitters at $z \sim 3$–4 with faint or currently absent HST counterparts.

At present, the Carousel Lens hosts 11 multiply imaged background sources with spectroscopic redshifts, summarized in Table~\ref{tab:sources-z} and shown in Fig.~\ref{fig:all-sources-jack}. Ten of these systems lie near the cluster core and display highly magnified, symmetric image configurations resembling galaxy-scale strong lenses. This morphology indicates a smooth and dynamically relaxed central mass distribution, with a relatively minor contribution from galaxy-scale substructures. Source~9 (and source~10, whose redshift remains unknown) lies in the vicinity of a massive subhalo, labeled $L_e$, located approximately $30''$ from the cluster center and provides constraints to this local substructure.

Behind the cluster core, there are eight distinct source planes with spectroscopic redshifts, spanning $0.96 \lesssim z \lesssim 4.1$, and strongly lensed by the same main cluster halo, which provide constraints on cosmological parameters through the redshift dependence of the lensing deflection ratios. In this work, we use the HST-visible sources 1–7, which cover the range of $0.96 \lesssim z \lesssim 1.6$ in four redshift planes to constrain cosmology, and additionally, we include source~9 to constrain the mass associated with subhalo $L_e$. 
We note that the deeper MUSE observations presented by Paper~I became available after the lens modeling was completed. Moreover, the higher-redshift sources currently lack high-quality high-resolution imaging: source~8 has only low signal-to-noise data, while sources~11, 12, and 13 have none. For these reasons, they are not included in the present analysis; however, we present a forecast of the cosmological constraints that could be achieved if these sources were incorporated with suitable imaging data.

\section{Lens Modeling} \label{sec:lens_model}

The Carousel Lens was first modeled in Paper~0 using two elliptical power-law (EPL) mass profiles plus external shear. Here, we apply a more comprehensive model with redshifts from additional lensed sources and additional mass components.

We model the Carousel lens using the \gigal framework \citep{Gu2022}, which provides a fully parametric description of the lens system. The \gigal software combines GPU-accelerated forward modeling of the lensed surface brightness with state-of-the-art, gradient-based inference techniques. All stages of the inference are gradient-informed through automatic differentiation implemented via the JAX Python package \citep{jax2018github}. 

The lens model comprises four mass components, an external shear component, and six extended sources. Each of the four mass components is parameterized with an EPL mass profile. Following the results of Paper~0, which demonstrated that the observed image configuration in the cluster core can be accurately reproduced with a relatively simple mass model, we adopt an explicit modeling of the dominant deflectors rather than a population-based treatment of cluster substructure.

\begin{figure}
    \centering
    \includegraphics[width=\linewidth]{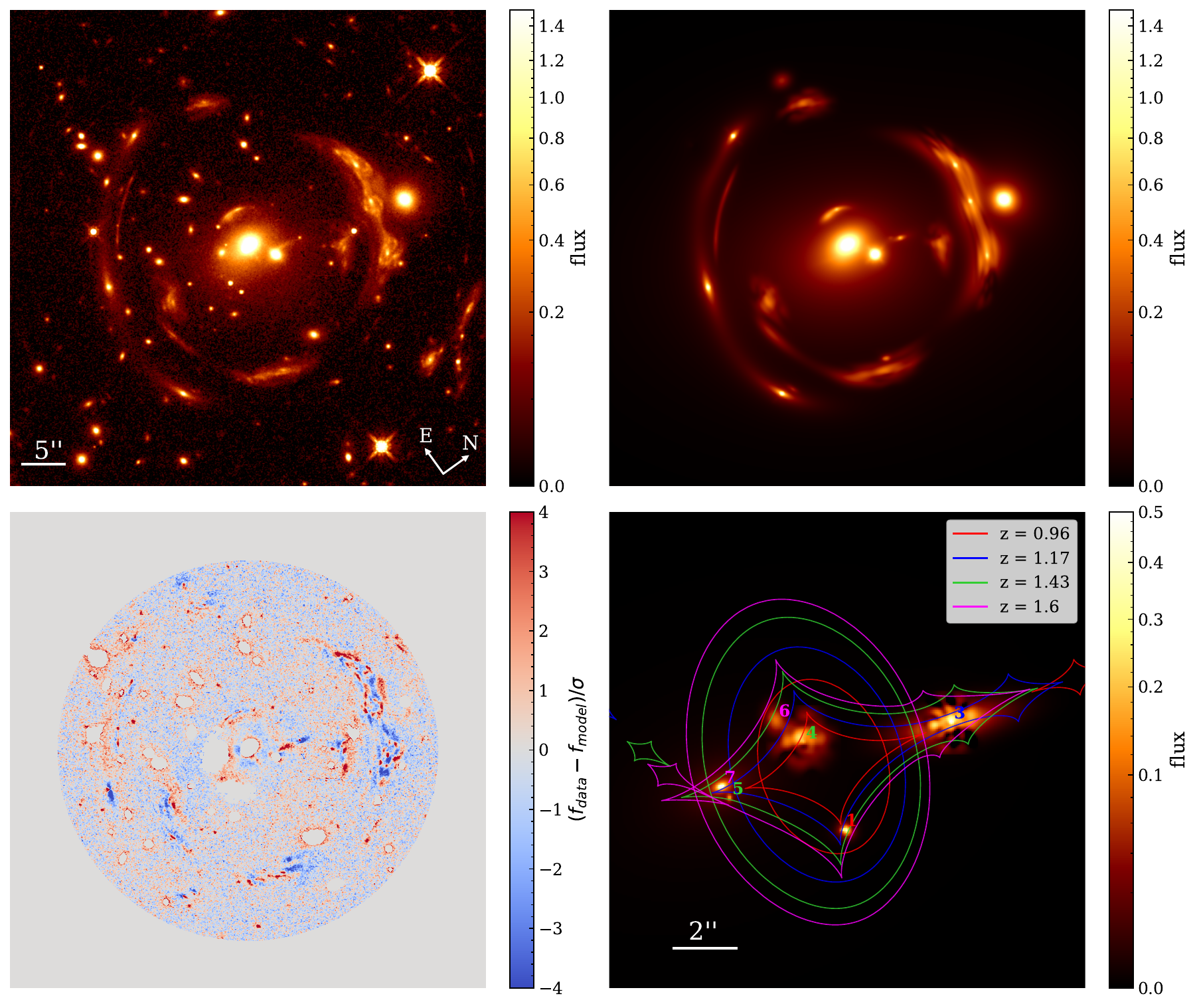}
    \caption{Strong lensing surface-brightness model. \textit{Top left}: $800 \times 800$ pixel HST/F140W cutout of the cluster core used as a constraint in the lens modeling. \textit{Top right}: Best-fit lensed surface-brightness model. \textit{Bottom left}: Noise-normalized residuals between the model and the data, with cluster member galaxies and non–strongly lensed objects masked. \textit{Bottom right}: Reconstructed unlensed source-plane surface-brightness, overlaid with the lens caustics for each source redshift; individual sources are labeled and color-coded consistently with the caustics.}
    \label{fig:model-residual}
\end{figure}

The main mass component has a free position around the central cluster members $L_a$ and $L_b$. The second component is assigned a prior on its position and ellipticity to match the centroid and shape of galaxy $L_d$; this clump is primarily constrained by image family 3. A third component is included with fixed position and ellipticity matching galaxy $L_e$ and is constrained mainly by image family 9. Finally, a fourth clump is introduced with a free position around galaxy $L_f$, where there is a high local density of cluster members, and is intended to capture the collective lensing effect of these nearby subhalos.

We note that this approach differs from the more commonly adopted treatment in which cluster substructure is modeled using scaling relations such as the Faber–Jackson relation. Our simplified parameterization is motivated by the ability of a small number of dominant components to reproduce the strong-lensing observables in this system, and we quantify the impact of this choice through dedicated simulations. The resulting systematic uncertainties associated with the subhalo treatment are discussed in Appendix~\ref{apx:subhalo-scatter}.

The model incorporates extended surface brightness profiles for sources 1, 3, 4, 5, 6, and 7, and is further constrained by the positions of source~9 images. Sources 5, 6, and 7 are modeled with a single S\'ersic profile, reproducing their positions, shapes, and extent. Source 1 is modeled with two S\'ersic components to capture its structure. The morphologies of sources 3 and 4 are more complex and require additional flexibility. Source 3 exhibits a distinct bulge, disk, and spiral arms; we model it with two S\'ersic profiles to represent the bulge and disk, and a shapelet basis of order 10 to reproduce the spiral arms and finer features \citep{Refregier2003a, Refregier2003b}. Source 4 is modeled with a S\'ersic profile combined with a shapelet basis of order 10, which captures its main structural features.

We also include the surface brightness distribution of cluster members $L_a$, $L_b$, $L_c$, and $L_d$, each modeled with a single S\'ersic profile.

Compared to the model in Paper~0, this one includes two additional mass components (EPLs) and two additional extended sources: sources 6 and 7. The model has a total of 237 free parameters: 22 related to the lens mass, 28 to the light from cluster members, 55 S\'ersic parameters for the sources, and 132 shapelet parameters. These parameters are constrained by approximately $3\times10^5$ pixels within an $800 \times 800$ HST cutout---after masking foreground galaxies and unmodeled cluster members---together with three additional positional constraints from source~9 located outside this central region.

The inference is performed by simulating the surface brightness of the lensed sources, $\mathcal{I}_{\text{model}}(x, y; \Theta)$, given the lens parameters $\Theta$, and comparing it pixel-by-pixel to the observed image, $\mathcal{I}_{\text{obs}}(x, y)$. The $\chi^2$ is computed as:

\begin{equation} \label{eq:chi2_pixels}
    \chi^2_{\text{pix}} = \sum_{x,y} \frac{\left( \mathcal{I}_{\text{obs}}(x, y) - \mathcal{I}_{\text{model}}(x, y; \Theta) \right)^2}{\sigma^2_{\text{pix}}(x, y; \Theta)}
\end{equation}
\noindent
where $\sigma^2_{\text{pix}}(x, y; \Theta)$ represents the image noise, including both background and Poisson noise. The model is further constrained using the positions of image family 9, with the $\chi^2$ computed in the source plane. In this case, the $\chi^2$ is defined by the distance between the mapped image positions in the source plane, $\bm{\beta}$, and their barycenter, $\langle \bm{\beta} \rangle$:
\begin{equation} \label{eq:chi2_sp_positions}
    \chi^2_{\beta} = \sum_{i=1}^{N} \sum_{j=1}^{n_i} \frac{\left( \bm{\beta}(\bm{\theta}_{\text{obs}, j}; \Theta) - \langle \bm{\beta}(\bm{\theta}_{\text{obs}, j}; \Theta) \rangle \right)^2}{\mu^{-2}(\bm{\theta}_{\text{obs},j}; \Theta) \; \sigma^2_{\theta, j}}
\end{equation}
for a system comprising $N$ sources, where each source~$i$ has $n_i$ images with positions $\bm{\theta}_j$ and astrometric uncertainties $\sigma^2_{\theta, j}$. The term $\mu^{-2}(\bm{\theta}_{\text{obs},j}; \Theta)$ corresponds to the inverse squared magnification at the image position $\bm{\theta}_{\text{obs},j}$, used to scale the astrometric error to the source plane.

The likelihood is given by the weighted sum of both terms:
\begin{align} \label{eq:likelihood}
    log \, \mathcal{L}_k &= -\frac{1}{2} \left[ \chi^2_k + log \left( 2 \pi 
 \sigma^2_k \right)\right] \\
    log \, \mathcal{L}_{\text{tot}} &= log \, \mathcal{L}_{\text{pix}} + q \, log \, \mathcal{L}_{\beta}
\end{align}
where $k$ represents the type of constraint (pixels or positions), and $q$ is a hyperparameter of the modeling process weighting the two likelihoods. 
Because the number of pixel constraints ($\sim 800^2$) vastly exceeds the few positional constraints (three images with two coordinates each), the pixel term would otherwise dominate the likelihood. We therefore adopt $q = 10$ to compensate for this imbalance and ensure that the positional constraints meaningfully contribute, while the overall fit remains primarily driven by the pixels.

The lens model was first obtained through multi-start gradient optimization \citep[the MAP stage of the \texttt{GIGA-Lens} pipeline; see][]{Gu2022}, using a cutout of $800 \times 800$ pixels from the HST-WFC3 F140W band image. We use the F140W band rather than the F200LP filter because the redder sources, in particular sources 3 and 7, are significantly fainter in F200LP. Modeling the $800 \times 800$-pixel cutout is memory-intensive due to the intermediate computations required, limiting the number of models that can be evaluated in parallel. To address this, we performed an initial exploration of the parameter space on a rebinned version of the image ($200 \times 200$ pixels, using a $4 \times 4$ binning), which significantly reduced memory usage and allowed us to evaluate $2000$ models in parallel. This strategy helped mitigate the risk of converging to local minima, which is particularly important given the high dimensionality of the model.

The optimization was run on a node with four A100 GPUs (80 GB each) at the NERSC \emph{Perlmutter} supercomputing facility. We began with a simplified configuration including only sources 4 and 6---both are quadruply imaged\footnote{Source 4 also has a fifth radial image very close to the cluster core identified in the MUSE data (Paper~I) but not visible in HST. It is not modeled in this work.}---while masking all other sources. Additional extended sources were incorporated progressively, with each stage initialized from $2000$ randomized models informed by the previous best-fitting solution until the full model complexity was reached. Finally, we refined the model using the full-resolution $800 \times 800$ cutout image, exploring 100 models initialized around the previously best-fitting solution. In the final stage, all 237 model parameters were simultaneously optimized using gradients, without relying on linear inversion of the source amplitudes. From this final model, we performed a Markov Chain Monte Carlo (MCMC) sampling to characterize the posterior distribution of the most relevant lens parameters and estimate their statistical uncertainties, following the \texttt{GIGA-Lens} pipeline \citep[the SVI and HMC stages; see][]{Gu2022}. Further details on the MCMC implementation and convergence diagnostics are provided in Section~\ref{sec:cosmo} and Appendix~\ref{apx:chains}. The full optimization stage requires approximately 100 minutes of GPU time, while the posterior sampling takes approximately 140 minutes. We note that these timings refer only to the final production runs; the overall modeling effort, including iterative refinement, validation, and convergence testing, required substantially longer development time. 
Nevertheless, this is remarkably fast for such a large system.

From this procedure, we obtained the model shown in Fig.~\ref{fig:model-residual}. The model reproduces the image configuration of all six families in the core region. While sources 1–5 had already been recovered by the model of Paper~0, our new model also accounts for an additional quadruple-image system (source~6) and a cusp with a radial arc (source~7), and also achieves a lower reduced $\chi^2$ than in Paper~0. The model further predicts a counterimage for source~5; however, the low magnification does not allow us to detect this image with our current observations with HST or MUSE.

\begin{table}[thbp]
    \centering    
    \caption{Lens model parameters.}
    \label{tab:model-parameters}
    \begin{tabular}{ccccccc}
        \hline
        \hline
         Component & RA & DEC & $\theta_E$ & $\gamma$ & $q$ & $\gamma_{\text{ext}}$ \\
         & deg & deg & $['']$ &  &  &  \\
         \hline
         Main halo & 90.985136 & -35.968265 & $12.738 \pm 0.005$ & $1.571 \pm 0.001$ & $0.750 \pm 0.001$ & - \\
         $L_d$ & 90.983299 & -35.963231 & $1.11$ & $2.42$ & $0.56$ & - \\
         $L_e$ & 90.994836 & -35.976243 & $1.68$ & $2.02$ & $0.85$ & - \\
         $L_f^*$ & 90.992539 & -35.970336 & $0.73$ & $2.22$ & $0.45$ & - \\
         External Shear & - & - & - & - & - & $0.0260 \pm 0.0003$ \\
         \hline
         \multicolumn{7}{L{16cm}}{\footnotesize{Here $\theta_E$ is the Einstein radius respect to $z_{ref} = 1.432$, $\gamma$ is the power-law slope, $q$ is the axis ratio and $\gamma_{ext}$ is the magnitude of the external shear. Only the main halo and external shear have statistical uncertainties, as other components are kept fixed during sampling (see Section~\ref{sec:cosmo}). The location of the four mass components can be found in Fig.~\ref{fig:critical-positions}.}} \\
         \multicolumn{7}{l}{\footnotesize{* This component represents the collective mass of cluster members around $L_f$.}} \\
    \end{tabular}
\end{table}

\begin{table}[thbp]
    \centering    
    \caption{Average RMS per image family. }
    \label{tab:rms}
    \begin{tabular}{ccc}
        \hline
        \hline
         Source & HST images$^a$ & RMS \\
          &  & $['']$ \\
         \hline
         1 & 2 & 0.11 \\
         3 & 3 & 0.16 \\
         4 & 4 & 0.43 \\
         5 & 3 & 0.15 \\
         6 & 4 & 0.74 \\
         7 & 4 & 0.09$^b$ \\
         9 & 3 & 0.34 \\
         \hline
        \multicolumn{3}{L{4cm}}{\footnotesize{$^a$ Number of images visible in the HST cutouts.}} \\
        \multicolumn{3}{l}{\footnotesize{$^b$ Excluding the radial arc 7d$^*$ (see text).}} \\
    \end{tabular}
\end{table}

\begin{figure}[tbph]
    \centering
    \includegraphics[width=\linewidth]{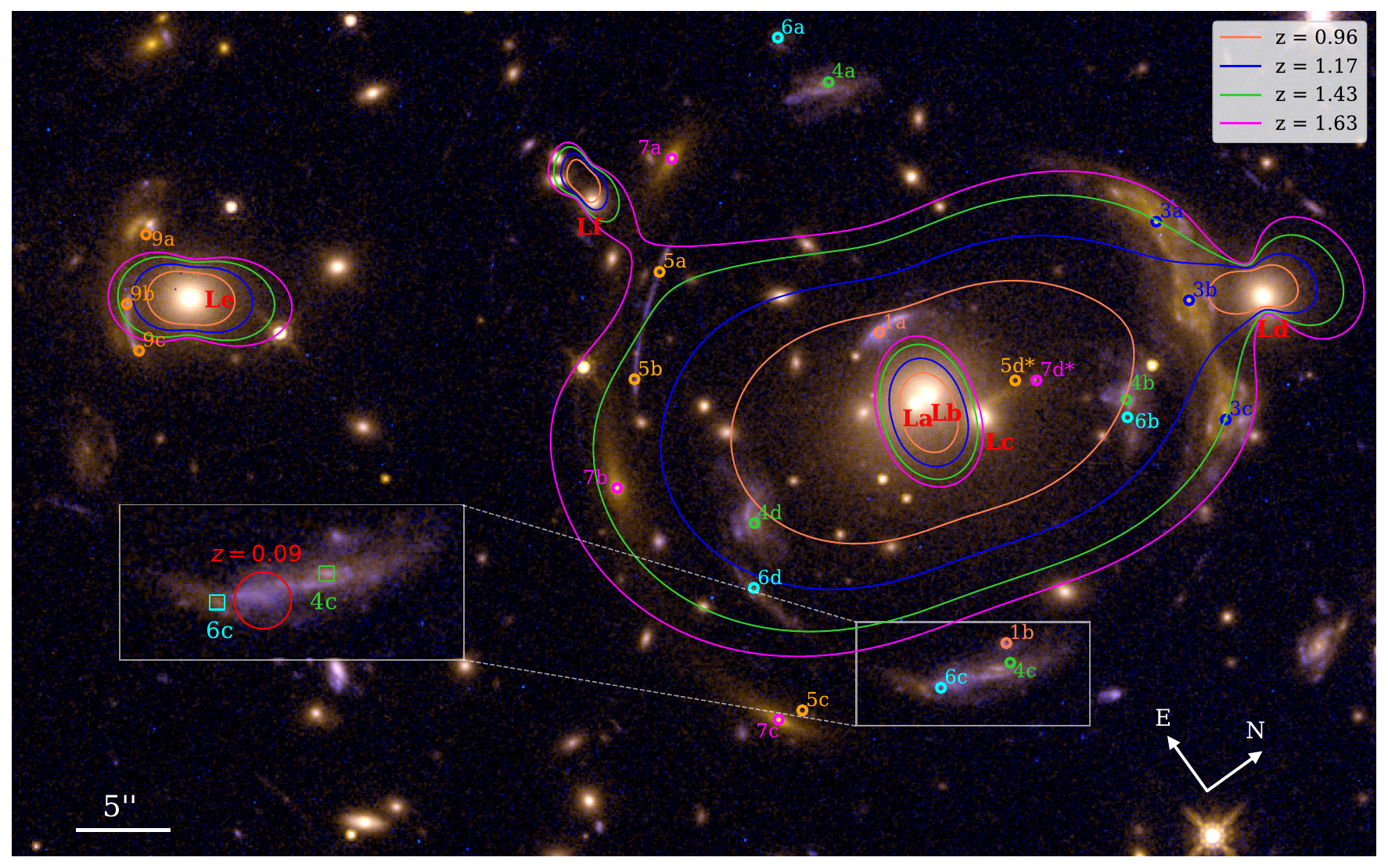}
    \caption{Critical curves and predicted source positions. The image is an RGB composite of the cluster core, constructed from HST observations, where the F140W filter is mapped to the red channel, the F200LP filter to the blue channel, and the average of the two filters to the green channel. Colored curves indicate the critical lines of the best-fit lens model for the different source planes. Colored circles mark the model-predicted positions of multiple images for each source. Cluster member galaxies included in the lens model are labeled in red. The inset shows a zoomed-in view of the region around images 4c and 6c, highlighting a foreground galaxy at $z = 0.09$ whose light was not modeled in this work (see Section~\ref{sec:discussion_lens_model}).}
    \label{fig:critical-positions}
\end{figure}

Within the mask, the reduced chi-squared is $\chi_{\nu}^2 = 1.38$. 
The most significant residuals occur for source~3, whose complex morphology cannot be captured with an order 10 shapelets model. Although a higher-order basis could improve the fit, it would require more memory than is available with our current computational resources. Nevertheless, the model successfully recovers the bulge position, as well as the overall shape and morphology of its three images. Residuals are also high near the radial image of source~7, which appears $1.5''$ farther from the core compared with the suspected, but not yet spectroscopically confirmed, radial image (7d$^*$ in Fig.~\ref{fig:critical-positions}), possibly indicating a flatter inner mass profile.

Fig.~\ref{fig:critical-positions} shows the critical curves and predicted image positions. Excluding the radial arc 7d$^*$, the RMS for all source families is $0.40''$. A breakdown of the RMS by image family is given in Table~\ref{tab:rms}. The total RMS is dominated by source~6, with the largest discrepancies found for images 6b and 6c, followed by 4c. We remark that images 4c and 6c are blended with a foreground galaxy (Paper~I), which was not recognized during modeling. This suggests that the higher offsets are due to blending between components rather than deficiencies in the mass model itself. When the two blended images and the radial arc of source~7 are excluded, the RMS for the remaining 20 images is $0.17''$, which is low compared to the typical values for cluster lenses $\sim 0.3-0.5''$ \citep[e.g.][]{Sharon2020, Caminha2022, Cerny2025}.

\begin{figure}[tbph]
    \centering
    \includegraphics[width=0.9\linewidth]{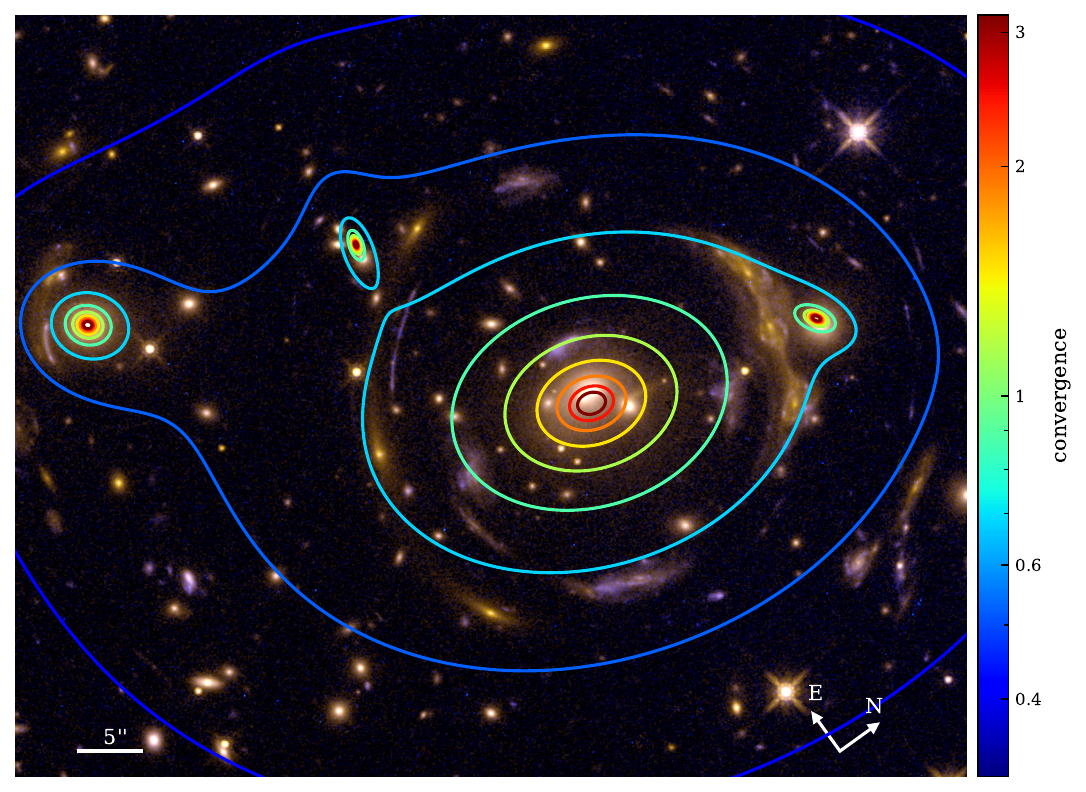}
    \caption{Convergence ($\kappa$) contours of the best-fit lens mass model overlaid on the RGB composite image of the cluster core.}
    \label{fig:convergence}
\end{figure}

Compared to the model from Paper~0, the Einstein radius of the main halo and $L_d$ are consistent; however, they differ in the slope, as the main halo is flatter in our model, while the mass of $L_d$ is steeper. The external shear is an order of magnitude lower than in the previous model, which is likely attributed to the inclusion of a new substructure around the bright galaxy $L_e$ in this new model. 

The convergence map of the lens model shown in Fig.~\ref{fig:convergence} indicates that the mass distribution in the vicinity of multiple-image systems 1–7 is dominated by the main cluster-scale halo, with a secondary contribution from the clump associated with $L_d$, in agreement with the results of Paper~0. The massive subhalo $L_e$, together with the less massive substructure centered on the member $L_f$ (together with nearby cluster members), exerts a more long-distance influence, contributing additional mass to the eastern region of the cluster core. This extra mass is required to reproduce the larger separation between images 7a and 7b compared to that between 7b and 7c (see also Fig.~\ref{fig:critical-positions}). Notably, the substructure associated with $L_f$, whose position and ellipticity are allowed to vary, is well aligned with the surrounding cluster members. In contrast, the $L_d$ clump exhibits a modest positional offset and a higher elongation relative to its luminous counterpart. This component is tightly constrained by image family 3, and the observed discrepancy may reflect local variations in the mass distribution---potentially due to additional nearby members---that are effectively absorbed by the $L_d$ parameterization. Overall, this difference remains small, and the high level of detail with which the model reproduces the lensing configuration of source~3 suggests that the mass distribution in this region is robustly constrained.

\section{Cosmological Constraints} \label{sec:cosmo}
In this section, we present cosmological constraints from the Carousel Lens.

\subsection{Methodology for cosmological inference}

In the model, we include an additional parameter, the deflection ratio $\eta_i$, for each image family, which accounts for the differing lensing deflection at each source plane. The deflection ratios are defined relative to a reference redshift $z_{\text{ref}} = 1.432$, corresponding to the redshift of sources 4 and 5. These parameters are treated as free during the model inference, except for those corresponding to sources at $z_{\text{ref}}$, for which $\eta = 1$ by definition.

The deflection ratios for sources 1, 3, 6, and 7, relative to sources 4 and 5, together with their spectroscopic redshifts, allow us to obtain cosmological constraints for each one. 

After obtaining the best-fit lens model through the multi-start gradient optimization (the MAP stage of \texttt{GIGA-Lens}) described in the previous section, we constrain the cosmological parameters through a two-step inference process. First, we explore the posterior distribution of the lens model using MCMC sampling, obtaining the posterior distribution of the deflection ratios $\eta_i$ (see Subsection~\ref{sec:mcmc}). We then use the mean values of $\eta_i$ and their covariance matrix $C_{ij}$---which are well approximated by a multivariate normal distribution, as shown in Fig.~\ref{fig:mcmc_cornerplot} of Appendix~\ref{apx:chains}---to explore the cosmological parameters $\Omega_m$ and $w$, which relate the model-derived deflection ratios to the corresponding source redshifts:
\begin{equation} \label{eq:chi2_cosmo}
    \chi^2_{\eta} = \sum_{i,j} \left( \eta_i - \eta(z_l, z_i, z_{\text{ref}}; \Omega_m, w) \right)
    C^{-1}_{ij}
    \left( \eta_j - \eta(z_l, z_j, z_{\text{ref}}; \Omega_m, w) \right)
\end{equation}
This second exploration step is performed by integrating the likelihood over a regular grid of cosmological parameters (see Subsection~\ref{sec:wCDM}).

\subsection{MCMC Sampling of the Lens Model} \label{sec:mcmc}
For the MCMC step, we use a Hamiltonian Monte Carlo (HMC) kernel, preconditioned with the covariance matrix obtained through Stochastic Variational Inference (SVI), which enables efficient sampling of highly correlated parameters \citep[see e.g.][]{Gu2022, Cikota2023, Urcelay2025, Huang2025b, Huang2026, baltasar2026}. In addition, we employ a parallel tempering scheme to enhance robustness and ensure efficient exploration of multimodal or otherwise complex posterior distributions. More details on the MCMC method and the posterior distribution are provided in Appendix~\ref{apx:chains}.

During this sampling, we fixed the majority of the lens model parameters, as sampling them all was computationally prohibitive. Instead, we allowed the most relevant parameters---namely the Einstein radius, EPL slope, and ellipticities of the main halo, the external shear, and the source positions---to vary freely, since these are expected to be the most correlated with the deflection ratios. Tests in which a small number of additional parameters were either included or fixed, including the source Sérsic amplitudes and the Einstein radius of $L_d$, confirmed that their impact on the final posterior was negligible. Furthermore, during the MCMC, we masked the radial image of source 7 (i.e., 7d$^*$) to avoid potential biases arising from its mismatch with the model. 

While Bayesian modeling is widely regarded as the gold standard for inference, it does not guarantee that the inferred uncertainties are fully reliable, and the model may still be subject to systematic effects. We therefore apply quantitative convergence tests and explicitly evaluate both statistical and systematic uncertainties, as described below.

\subsection{Calibration of statistical uncertainties}

When performing pixel-level strong-lensing modeling, model misspecification (i.e., the true data-generating process lying outside the assumed model parameter space) is unavoidable. The parametric mass distribution adopted in the model is likely a simplified representation of the true underlying mass distribution. This limitation becomes increasingly relevant when fitting high-resolution surface-brightness data, where the large number of pixel constraints and the high signal-to-noise ratio (S/N) make even small discrepancies between the model and the data statistically significant.

In our case, the lens model reproduces the observed image positions and morphologies and successfully predicts independent quantities (see Section~\ref{sec:lens_model} and Subsection~\ref{sec:discussion_lens_model}). However, as indicated by a reduced $\chi^2_\nu > 1$ for the high-S/N pixels, the model does not reproduce all surface-brightness details within the nominal per-pixel uncertainties. Under a Gaussian likelihood assumption (equations~\eqref{eq:chi2_pixels} and \eqref{eq:likelihood}), this implies that some regions of the data formally lie outside the expected fluctuations of the model.

If not accounted for, such model–data mismatch can lead to underestimated statistical uncertainties. In Bayesian inference, when the assumed model does not perfectly describe the true data-generating process, the posterior distribution may still concentrate near the best-fitting parameter values, but its width can be incorrectly estimated (e.g., \citealt{Kleijn2012}; also discussed in an astrophysical context by \citealt{RomeroShaw2022}, and noted by \citealt{Hogg2010}). In practice, this means that the posterior can appear artificially tight even when the model does not fully reproduce the data.

A common and pragmatic approach in astrophysical modeling is to introduce an ``error floor'' or rescale the data uncertainties so that the reduced $\chi^2_\nu$ is approximately unity, a procedure widely used in photometric and spectroscopic analyses to account for underestimated noise or residual systematics (e.g., \citealt{huang2017, Scolnic2018}). More generally, in the presence of model misspecification, several authors have advocated tempering or rescaling the likelihood during the MCMC to obtain more reliable uncertainty estimates (e.g., \citealt{Grunwald2014}; \citealt{Thomas2019}). Following this logic, we adopt a conservative calibration of the statistical uncertainties during the MCMC step, i.e., one that would lead to larger uncertainties. Specifically, we rescale the total log-likelihood by a factor of $1/\hat{\chi}^2_\nu$, where $\hat{\chi}^2_\nu = 2.20$ is computed from pixels with ${\rm SNR} > 2$, which are most sensitive to small model discrepancies. This procedure is equivalent to inflating the per-pixel uncertainties $\sigma_{\rm pix}$ by a factor of $\sqrt{\hat{\chi}^2_\nu}$, ensuring that the high-S/N residuals are statistically consistent with the assumed Gaussian noise model.

Because our priors are uniform (with the exception of the main halo axis ratio, which is Gaussian following \citet{Gu2022} but remains broad compared to the posterior), this rescaling does not shift the maximum-likelihood solution. Instead, it broadens the posterior distribution, allowing a more realistic exploration of parameter space and yielding conservative estimates of the statistical uncertainties, $\sigma_{\rm stat}$. We therefore refer to this step as a calibration of the statistical uncertainties, reflecting the intrinsic model–data mismatch while preserving the predictive power and overall quality of the lens model.

The calibrated posterior constraints from the MCMC are summarized below. The mean values and $1\sigma$ confidence intervals of the lens parameters are listed in Table~\ref{tab:model-parameters}, while the corresponding deflection ratios are reported in Table~\ref{tab:deflection-ratios}, together with their estimated systematic uncertainties (see the following section). The full posterior distribution is provided in Appendix~\ref{apx:chains}.

\begin{table}[htb]
    \centering
    \hspace{1cm}\caption{Deflection ratios mean and uncertainties.}
    \begin{tabular}{ccccc}
        \hline
        \hline
         Source & redshift & $\eta$ & $\sigma_{stat}$ & $\sigma_{sys}$ \\
         \hline
         1 & 0.962 & $0.7511$ & $0.0007$ & $0.0057$ \\
         3 & 1.166 & $0.8868$ & $0.0004$ & $0.0036$ \\
         4 & 1.432 & 1        & -        & - \\
         5 & 1.432 & 1        & -        & - \\
         6 & 1.656 & $1.0627$ & $0.0009$ & $0.0022$ \\
         7 & 1.627 & $1.0598$ & $0.0003$ & $0.0025$ \\
         \hline
         \multicolumn{5}{L{6cm}}{\footnotesize{Here $\eta$ corresponds to the mean deflection ratio of each source relative to $z_{\text{ref}} = 1.432$. $\sigma_{stat}$ is the $1\sigma$ statistical uncertainty of $\eta$, and $\sigma_{sys}$ the systematic one, which are two to nine times larger than $\sigma_{stat}$.}}
    \end{tabular}
    \label{tab:deflection-ratios}
\end{table}

\subsection{Estimation of systematic uncertainties}

There are likely systematic uncertainties that affect our cosmological constraints. Relevant sources of systematic uncertainty in our inference include: (i) the fidelity of the lens mass model (e.g., elliptical power-law versus NFW+baryons); (ii) unmodeled mass substructures, such as cluster members; (iii) multiplane lensing effects; and (iv) the ability of the adopted source surface-brightness parametrization to reproduce the observed image structure. These can be addressed by increasing the model complexity and testing various model assumptions, which is work currently underway.
In this work, we perform dedicated simulations to estimate the impact of (ii), which in our view likely represent the dominant systematic effect for cosmological inference.
We further test the impact of (iii).

Regarding the source modeling, the adopted surface-brightness parametrization should ideally be flexible enough to reproduce the fine morphological details of the lensed images. In practice, we find that for each family at least one image is reproduced with almost noise-like residuals (see Fig.~\ref{fig:model-residual}), suggesting that remaining structured residuals are more likely due to mass-model limitations (resulting in a small discrepancy in the centroids of the other images) rather than source modeling. The main exception is source~3, for which the residuals indicate that a more complex source parametrization may be required. We have not explicitly quantified the systematic uncertainty associated with the choice of source parametrization through dedicated simulations; this will be explored in future work. Nevertheless, while detailed source structure helps constrain the lens model, the deflection ratios are primarily driven by image positions, and the inclusion of extended surface-brightness information mainly reduces statistical uncertainties. We therefore do not expect the adopted source parametrization to introduce a significant bias in the inferred cosmological parameters.

To assess the impact of the remaining dominant systematics, we rely on dedicated simulations. We simulated 10 mock systems designed to reproduce the configuration of the Carousel Lens, but with higher complexity than our fiducial model (more details are provided in Appendix~\ref{apx:subhalo-scatter}). In each system, the main halo is simulated with an NFW profile, while the cluster members are represented by dual pseudo-isothermal elliptical (dPIE) subhalos, as defined by \citealt{Eliasdottir2007} and widely adopted in cluster strong-lensing analyses (e.g., \citealt{Sharon2020, Caminha2022, Cerny2025}). The NFW characteristic density is set to reproduce the observed Einstein radius of source 4, while the concentration is drawn from the mean mass–concentration–redshift relation of \citet{Diemer2019} for a halo of mass $M_{200c} = 10^{15}M_\odot$ at $z = 0.49$. The characteristic density and scale radius of the NFW are kept fixed across all simulations, introducing a potential bias but no additional scatter. The subhalo properties are set through luminosity-based scaling relations, with velocity dispersion $\sigma_{LT}$ and truncation radius $r_{cut}$ given by
\begin{equation}
\sigma_{LT} = \sigma_{LT}^* \left( \frac{L}{L^*} \right)^{1/4}, \quad
r_{cut} = r_{cut}^* \left( \frac{L}{L^*} \right)^{1/2}
\end{equation}
where $\sigma_{LT}^*$ and $r_{cut}^*$ correspond to the values for a galaxy with luminosity $L^*$. We adopt the parameters $\sigma_{LT}^*$, $r_{cut}^*$, and $L^*$ from \citet{Bergamini2019}, who constrained them by combining strong lensing and stellar kinematics in three clusters. To account for intrinsic scatter, we introduce a 15\% random variation in both $\sigma_{LT}$ and $r_{cut}$ around the scaling relations, following assumptions similar to those adopted by \citet{DAloisio2011} and \citet{Bergamini2021}. Each of the 10 realizations corresponds to an independent random draw. All simulations assume a flat $\Lambda$CDM cosmology with $\Omega_m = 0.3$ and use the same source parameters as in our best-fit model.

We then modeled each simulated system using our fiducial model with four EPL components and quantified the deviation in the deflection ratio between the fitted model and the input simulation, $\Delta \eta$. For each source, we take the RMS of $\Delta \eta$ between the 10 simulations as an estimate of the systematic uncertainty $\sigma_{sys}$. This accounts both for differences in the mass profile of the main halo and for the uncertainties introduced by mass substructures. 

The resulting systematic uncertainties are reported in Table~\ref{tab:deflection-ratios}. We find that $\sigma_{\rm sys}$ is roughly 5 times larger than the statistical uncertainty, confirming that our constraints are dominated by systematics. The largest systematic uncertainty arises for source~1, which we attribute to the difference in the inner density slope between the EPL and NFW profiles. In Appendix~\ref{apx:multiplane}, we further assess the impact of multiple lens planes, but find it to be negligible compared to the other sources of error. 

We regard this estimate as an upper bound on the systematic effect due to cluster substructure. With the MUSE data (Paper~I), we can directly estimate the mass of each cluster member instead of relying on an average mass-to-light ratio with scatter. In addition, the models presented in Paper~0 and in this work indicate that the cluster is indeed quite relaxed, so the impact of substructure is likely to be small. Finally, in future work, we will systematically test the mass-profile assumptions for the main halo and the other three mass components and determine their effects on cosmological parameters.

\subsection{Constraints on $w$CDM} \label{sec:wCDM}
We use the mean values and associated uncertainties of the deflection ratios listed in Table~\ref{tab:deflection-ratios} to constrain a $w$CDM cosmology, following equation~\eqref{eq:chi2_cosmo}. To explore the $\Omega_m$–$w$ space, we integrate the likelihood over a regular grid rather than a second MCMC. This approach is more precise when the grid is sufficiently fine, given the low number of dimensions. We explore a $400 \times 400$ grid over a uniform prior.

We use uniform priors for both parameters to reflect physically motivated bounds:
\begin{itemize}
    \item $\Omega_m \sim \mathcal{U}(0, 1)$
    \item $w \sim \mathcal{U}(-2, -1/3)$
\end{itemize}
These ranges ensure consistency with a flat universe containing matter ($\Omega_m > 0$), and with accelerated expansion, which requires $w < -1/3$ and $\Omega_m < 1$. The lower bound on $w$ also avoids the regime of phantom dark energy ($w < -2$), which is typically considered unphysical in most cosmological models.



\begin{figure}[tbhp]
    \centering
    \captionsetup[subfigure]{oneside,margin={0.6cm, 0cm}}
    \subfloat[]{%
        \includegraphics[width=0.4\linewidth]{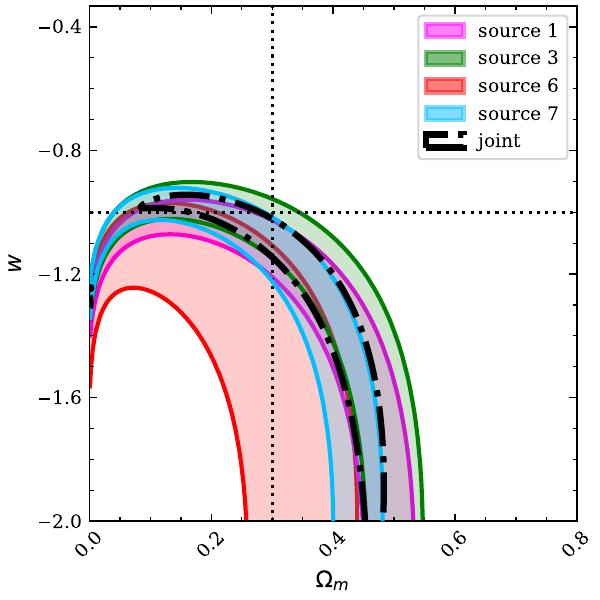}
        \label{fig:wCDM-single-tempered}%
    }
    \subfloat[]{%
        \includegraphics[width=0.4\linewidth]{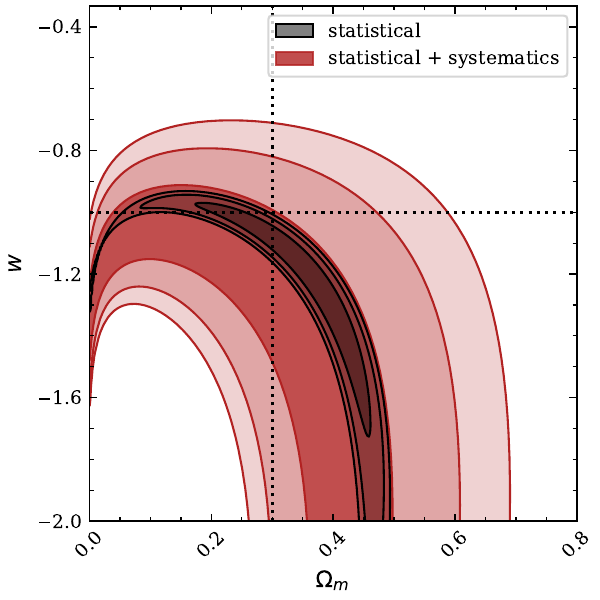}
        \label{fig:wCDM-systematics}%
    }
    \caption{$w$CDM constraints from the Carousel Lens under different uncertainty treatments. Panels: (a) 95\% probability contours from source 1 (pink), source 3 (green), source 6 (red), source 7 (sky blue), and their joint constraints (black), considering only statistical uncertainties;
    (b) Comparison of joint constraints with statistical and statistical plus systematic uncertainties, contours are 68, 95, and 99\% probability levels; the dashed line marks the concordance model ($\Omega_m=0.3$, $w=-1$).
    }
\end{figure}

Fig.~\ref{fig:wCDM-single-tempered} shows the individual and joint constraints on $\Omega_m$ and $w$ for these four sources, considering statistical uncertainties only.

The combined constraints, incorporating both the statistical uncertainties and the systematic uncertainties ($\sigma_{\rm tot} = \sqrt{\sigma_{\rm stat}^2 + \sigma_{\rm sys}^2}$), are presented in Fig.~\ref{fig:wCDM-systematics}. These results are comparable to those obtained for individual clusters in \citet{Caminha2022}, as well as for the group lens SDSS J0100+1818 studied by \citet{Bolamperti2024}, which features three source planes. Notably, however, none of these previous works account for the additional systematics arising from the intrinsic scatter on the adopted scaling relations for group/cluster members.

The Carousel Lens alone yields $\Omega_m = 0.34^{+0.16}_{-0.13}$ and $w = -1.31^{+0.35}_{-0.32}$, which are consistent with $\Lambda$CDM but not yet as constraining as those from established probes. Nevertheless, the degeneracy direction of the Carousel Lens constraints is nearly orthogonal to that of the CMB, highlighting their strong complementarity, and comparable in orientation to that of SNe~Ia, enhancing the statistical power, while being subject to different systematics from either probe. As a result, combining the current HST-based constraints from the Carousel Lens with Planck 2018 \citep{planck2018} yields $\Omega_m = 0.26 \pm 0.03$ and $w = -1.2 \pm 0.1$, compared to $\Omega_m = 0.34 \pm 0.01$ and $w = -0.93 \pm 0.05$ obtained from the joint CMB and Union3 SNe~Ia sample \citep{Rubin2025}, with 2000 SNe~Ia.

\subsection{Constraints on $w_0w_a$CDM}

We also consider an evolving dark energy model parameterized by the commonly adopted $w_0w_a$CDM form \citep{Chevallier2001,Linder2003}, in which the equation of state evolves as
\begin{equation}
    w(z) = w_0 + w_a \frac{z}{1 + z}
\end{equation}
\noindent
where $w_0$ is the value at present time, and $w_a$ sets the time evolution. 

We constrain this model using the same two-step inference framework adopted for $w$CDM. Specifically, we use the mean deflection ratios and their uncertainties from Table~\ref{tab:deflection-ratios} to construct the likelihood in the $(\Omega_m, \, w_0, \, w_a)$ parameter space, integrating the likelihood resulting from equation~\eqref{eq:chi2_cosmo} over a $400 \times 400 \times 400$ grid. We adopt the same priors on $\Omega_m$ and $w_0$ as in Subsection~\ref{sec:wCDM}, and assume a uniform prior on $w_a$, $w_a \sim \mathcal{U}(-3, 1)$. We additionally impose the constraint $w_a < -w_0$ to ensure consistency with the existence of matter- and radiation-dominated eras, which guarantees that $w(z) < 0$ at early times.

\begin{figure}[tbhp]
    \centering
    \includegraphics[width=0.8\linewidth]{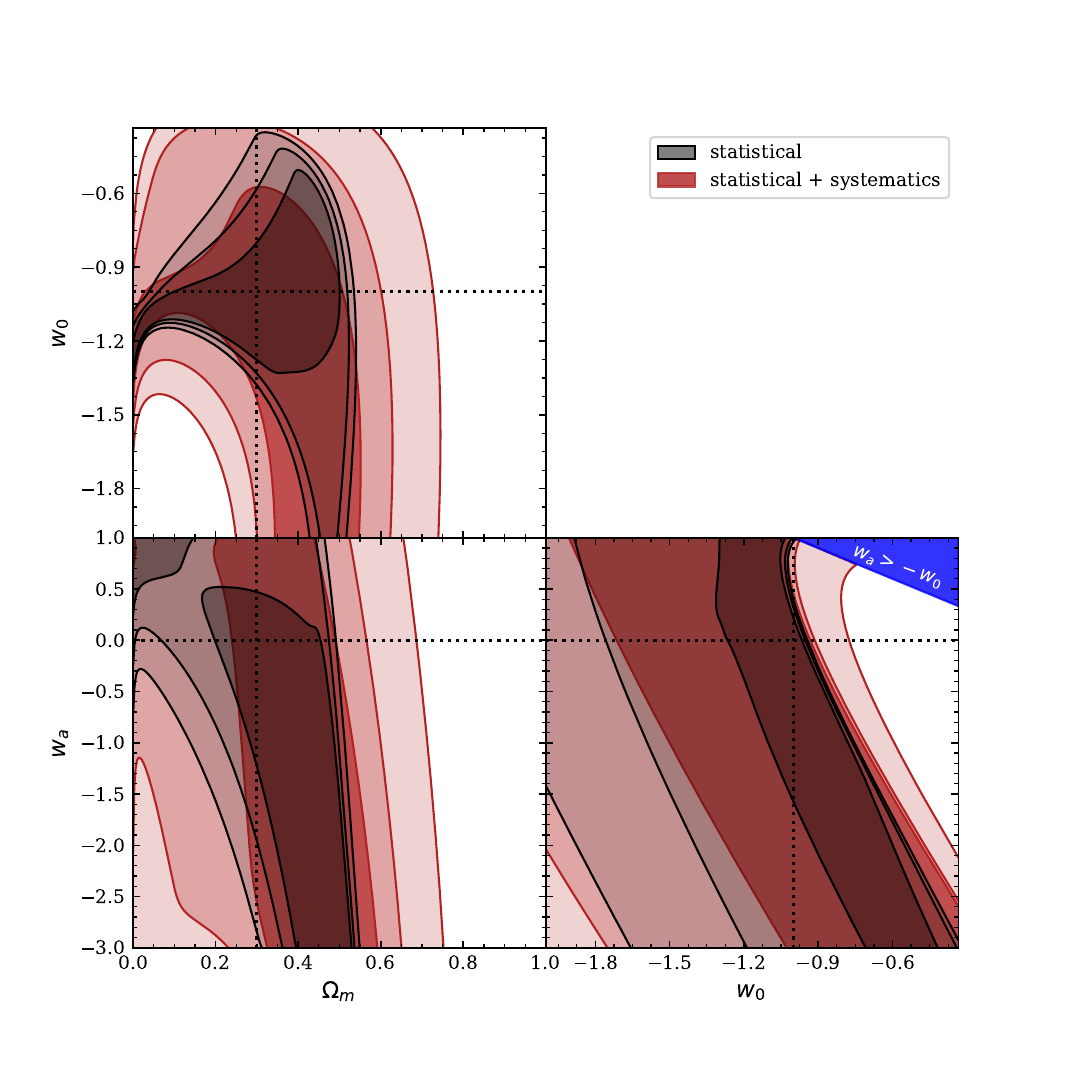}
    \caption{$w0wa$CDM constraints from the Carousel Lens alone under different uncertainty treatments. The corner plot shows comparison of joint constraints with statistical and statistical plus systematic uncertainties, with the contours being at 68, 95, and 99\% probability levels. The dashed line marks the concordance model ($\Omega_m=0.3$, $w_0=-1$, $w_a = 0$), and the blue region $w_a>-w_0$ is excluded as non-physical because it violates early matter domination. The shape of the contours differs between the two cases because the systematic uncertainties vary across sources and therefore do not correspond to a uniform rescaling of the statistical errors.
    }
    \label{fig:w0waCDM-systematics}
\end{figure}

The resulting constraints are shown in Fig.~\ref{fig:w0waCDM-systematics}. As expected from the limited redshift leverage of the current HST sources ($0.96 \lesssim z \lesssim 1.66$), the posterior exhibits a strong degeneracy between $w_0$ and $w_a$; however, $\Omega_m$ remains constrained within this parametrization. The marginalized 1D posteriors, including both statistical and systematic uncertainties, are $\Omega_m = 0.38^{+0.16}_{-0.13}$, $w_0 = -1.19^{+0.43}_{-0.53}$, and a nearly flat posterior for $w_a$.

\section{Discussion} \label{sec:discussion}
\subsection{Lens model improvements and limitations} \label{sec:discussion_lens_model}

Compared to our previous work (Paper~0), the new model provides several improvements. It successfully reproduces the observed image positions and morphologies (with the exception of the radial arc of source~7), achieves a lower RMS and $\chi^2_\nu$, and predicts previously unreported counter-images for source~6 (6a and 6b in this work; the naming differs slightly from Paper~0). In addition, the critical curve at $z=1.14$ now passes through the gap between the images of source 5, and the additional clump near $L_f$ aligns with the position and orientation of the sub-group of cluster members, increasing the physical consistency of the mass distribution.

Furthermore, when we started modeling the system, we had not yet secured a redshift for sources 6 and 7; however, our early model predicted $z \sim 1.6$ for both, close to the spectroscopic measurements from Paper~I, $z_6 = 1.656$ and $z_7 = 1.627$, respectively. In contrast, Paper~0 inferred a much higher redshift for source 7 ($z \sim 4.5$), driven by the steeper slope, lower elongation, and absence of secondary mass components in their model, which required a larger redshift to reproduce the observed image separation.

Despite these improvements, some limitations remain. In particular, the logarithmic slope of the main halo may be artificially constrained when combining lensing constraints that probe different radial regimes. Our current model adopts a constant logarithmic slope of $\gamma \sim 1.6$, whereas an NFW profile features a radially varying slope, transitioning from $\gamma = 1$ in the inner regions to $\gamma = 3$ at large radii. As a result, inner constraints may favor shallower density profiles, while outer constraints tend to prefer steeper ones. Because the EPL parameterization enforces a single power-law slope across all radii, it may lack sufficient flexibility to accommodate this behavior, potentially leading to over-constraining and systematic biases. 

Nevertheless, the inference itself is statistically robust within the adopted modeling framework. The convergence diagnostics indicate good mixing, with a potential scale reduction factor $\hat{R} \simeq 1$, well below the threshold of $\hat{R} < 1.1$ recommended by \citet{Gelman2014} (more details on the MCMC convergence are provided in Appendix~\ref{apx:chains}). This demonstrates that the posterior is well explored and that the inferred uncertainties are reliable given the assumed parameterization. 

The present parametrization follows that adopted in Paper~0, which showed that it reproduces the observed image positions and configurations in the cluster core with high accuracy using a relatively simple mass model. In this work, we therefore focus on maintaining a consistent lens model while assessing its impact on cosmological inference. Although this approach is not intended to represent the most physically motivated description of a cluster-scale mass distribution, our model still reproduces the strong-lensing observables present in the HST imaging. We quantify the associated systematic uncertainties through simulations, while a more general treatment of the cluster radial profile, the explicit inclusion of cluster members, and additional lensed sources is left to future work.

Finally, during the final stages of preparing this manuscript, we identified a foreground galaxy at $z = 0.09$ located between images 4c and 6c (see the inset in Fig.~\ref{fig:critical-positions}). This object may affect the lens model, particularly for source~6, whose inferred position shifts toward it, potentially explaining the increased tension and larger uncertainty of this source in the $w$CDM constraints. We note that this galaxy is relatively faint (absolute magnitude $M_{\rm F140W} \approx -17\,\rm AB\,mag$), and we therefore expect its lensing contribution to be minor; the dominant systematic effect is more likely due to contamination from its light. Properly accounting for this effect would require masking the affected image and/or incorporating multiband data to deblend the components, which would necessitate rebuilding the lens model and re-running the inference; we therefore defer this correction to future work. We nevertheless expect the impact on the cosmological results to be minor, as the effect is localized and the constraints rely on the full set of images within each family, with other images from sources~4 and~6 well reproduced.

\subsection{Image configuration and cosmological constraints}

The ability to constrain cosmological parameters from strong lensing depends not only on the number of sources at different redshifts, but also on the specific image configurations. In particular, certain configurations can break the degeneracy between the lens mass profile slope and cosmology (see Appendix~\ref{apx:geometry}), which could be the limiting factor for the statistical uncertainties \citep{Sharma2023}.

The Carousel Lens includes several such configurations. Source 1 forms a double system with asymmetric sensitivity to the mass slope, which helps decouple its effects from those of cosmology. Source 7 produces a radial arc, further constraining the local slope. Additionally, sources 4 and 5 lie at the same redshift but at different positions in the image plane, providing slope sensitivity without introducing cosmological dependence. As a result, the deflection ratios of different sources exhibit distinct correlations with the density slope (see the second column of Fig.~\ref{fig:mcmc_cornerplot} and Fig.~\ref{fig:geometry-cosmo} in the Appendix). This unique combination of features significantly reduces the slope–cosmology degeneracy, enabling tighter cosmological constraints. This richness in image configurations would be further enhanced with the inclusion of the newly identified MUSE sources and infrared imaging, as sources~4, 12, and 13 all exhibit radial images that provide additional leverage on the inner density slope of the mass profile.

\subsection{Comparison with other multi-source plane lenses}

The Carousel Lens provides an intermediate case between galaxy-scale and unrelaxed cluster-scale multi-source plane lenses. Galaxy-scale systems, such as those studied by \citet{Collett2014}, \citet{Sahu2025}, and \citet{Bowden2025}, typically have smaller Einstein radii with only two source planes (in case more source planes exists, they would be strongly affected by multi-lens plane effects), resulting in constraints on the mass profile slope at fewer locations. 
While galaxy lenses are more numerous and large populations of double source plane lenses could improve cosmological constraints \citep{Sharma2023}, individual systems suffer from limited image multiplicity and higher fractional uncertainties in the deflection ratios. Our constraints for the Carousel Lens are significantly stronger than those reported for galaxy-scale lenses by \citet{Collett2014}, \citet{Sahu2025}, and \citet{Bowden2025}, even when taking systematics into account.

Massive clusters can host more than a hundred multiply imaged background sources, but they are often dynamically unrelaxed and complex, typically requiring multiple massive clumps for accurate modeling. Different sources probe different regions of the cluster, which can increase systematics due to model dependence at many localized points. In contrast, the Carousel Lens is well described by a single massive halo plus a few galaxy-scale subhalos, with all the multiply imaged systems located within this region, similar to the group lens SDSS J0100+1818 \citep{Bolamperti2024}. This relatively simple mass distribution allows strong cosmological constraints despite a smaller number of sources, consistent with findings by \citet{Bolamperti2024}.

Furthermore, the larger Einstein radius, compared to galaxy-scale lenses, reduces the fractional uncertainty in deflection ratios and diminishes the impact of multi–lens-plane effects. While modeling of massive clusters often requires a large number of components---from cluster-scale halos down to individual members---the resulting uncertainties from unconstrained subhalos can dominate the uncertainty budget. Our analysis shows that, even when accounting for subhalo scatter, the Carousel Lens achieves constraints comparable to those of individual clusters studied by \citet{Caminha2022} and the SDSS J0100+1818 group \citep{Bolamperti2024}, neither of which included systematic effects, highlighting the advantages of relaxed systems with relatively simple mass distributions for multi-source plane cosmography. Moreover, when considering statistical uncertainties only---i.e., excluding the additional scatter associated with subhalos---our constraints are comparable to the joint constraints obtained by combining five clusters in \citet{Caminha2022}. We note that, while \citet{Caminha2022} explicitly models cluster subhalos using scaling relations, they do not include intrinsic scatter in those relations, limiting the comparison to statistical uncertainties only.

Finally, while surface-brightness (i.e., pixel-level) modeling is now standard for galaxy-scale lenses, its application to cluster-scale lenses remains relatively recent and computationally demanding \citep{Acebron2024}. In this work, we present the first cluster lens modeled at the surface-brightness level and used for multi–source-plane cosmography. The resolved source structure provides additional constraints on the deflection field, while the extended image morphology constrains higher-order lensing quantities such as magnification and flexion. Together, these constraints improve the precision of the lens model and, consequently, of the inferred cosmological parameters, consistent with previous findings that pixel-level modeling enhances the precision of $H_0$ cosmography \citep{Xie2025}. In future work, we aim to explore more flexible source parameterizations---particularly for source~3---to further assess the impact of source modeling assumptions on the inferred lens and cosmological parameters.

\begin{figure}
    \centering
    \captionsetup[subfigure]{oneside,margin={0.6cm, 0cm}}
    \subfloat[]{
        \includegraphics[height=0.32\linewidth]{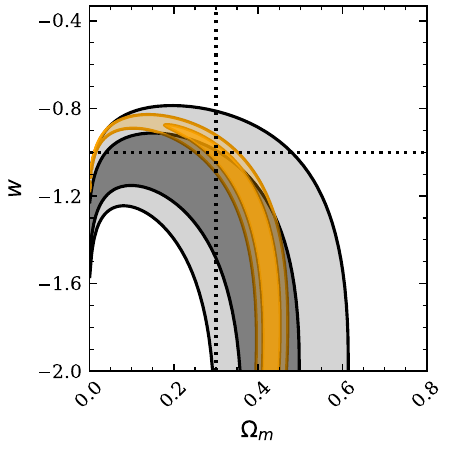}
        \label{fig:wCDM-sys_hst_muse}%
        }
    \subfloat[]{%
        \includegraphics[height=0.32\linewidth]{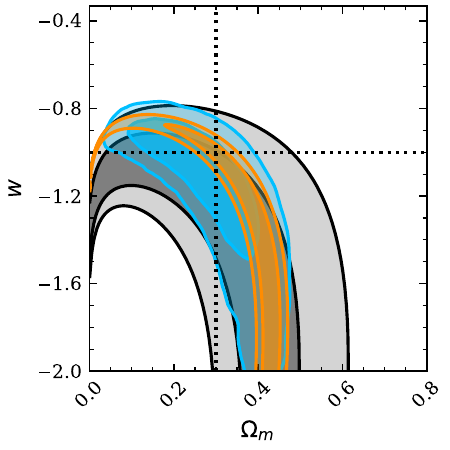}
        \label{fig:wCDM-sys_muse_caminha}%
    }
    \subfloat[]{%
        \includegraphics[height=0.32\linewidth]{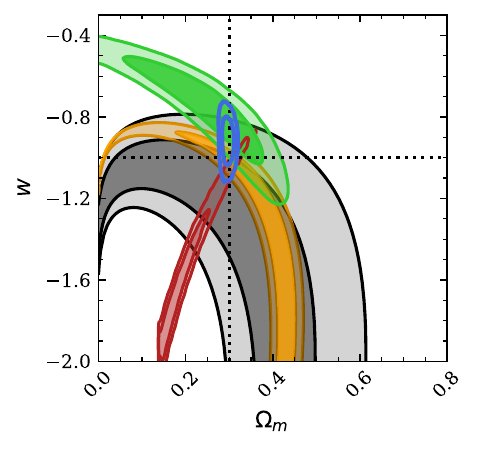}
        \label{fig:wCDM-sys_sn1a_cmb_bao}%
    } \\
    \includegraphics[width=\linewidth]{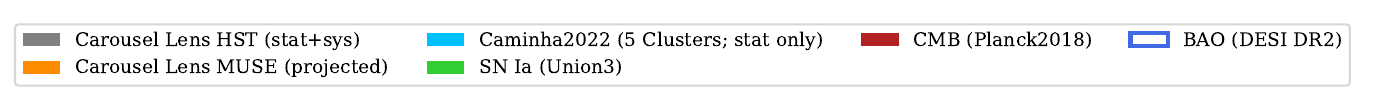}
    \caption{Predicted $w$CDM constraints from the Carousel Lens with additional sources. (a) Probability contours from current sources 1-7 from the HST data (gray), and predicted constraints when including sources 8, 11, and 12 from the MUSE data (orange)
    (b) HST (current) and MUSE (predicted) constraints, in both cases with both statistical and systematic uncertainties included, compared to \citet{Caminha2022}, which only includes statistical uncertainties. (c) Same constraints compared with CMB \citep{planck2018}, SNe Ia \citep{Rubin2025}, and BAO \citep{Abdul2025}. Contours  show 68 and 95\% probability levels. The predicted constraints assume $\Lambda$CDM with $\Omega_m = 0.3$, plus a random scatter and the average uncertainties of the HST $\eta$ measurements, including systematics.}
\end{figure}

\subsection{Projected $w$CDM constraints and comparison with other probes}

The constraining power of multiple-source-plane strong lenses increases when the sources span a wider redshift range. In our current analysis, the constraints are limited to five source planes that lie relatively close to each other in redshift: $z = 0.962$, $1.166$, $1.432$, $1.627$, and $1.656$. Deeper MUSE observations revealed four additional multiply imaged sources---sources 8, 11, 12, and 13 in Table~\ref{tab:sources-z}---spanning three distinct redshifts in the range $3 \lesssim z \lesssim 4$, with sources 12 and 13 lying at the same redshift. Incorporating these systems into the model will require high-resolution imaging of them, as well as the use of multi-band constraints, and will be the subject of future work. If the fractional uncertainties in the deflection ratios of these sources remain comparable to those of sources 1–7 ($0.4\%$, accounting for systematics)---a reasonable assumption given their observed image configurations---and if multi–lens-plane effects do not introduce significant additional systematics at these redshifts, as is generally the case for cluster-scale strong lenses, then the inclusion of higher-redshift sources has the potential to substantially tighten the cosmological constraints.

To illustrate this, we simulate the deflection ratios for the new sources under a flat $\Lambda$CDM cosmology with $\Omega_m = 0.3$, adopting the average scatter and uncertainty $\langle \sigma_{\rm tot} \rangle$ from sources 1–7. The predicted posterior distribution is shown in Fig.~\ref{fig:wCDM-sys_hst_muse}. Notably, the predicted constraints obtained by including the new MUSE-identified systems, with systematics taken into account, closely match those derived using only the HST sources (1–7) and statistical uncertainties alone (Fig.~\ref{fig:wCDM-single-tempered}). The additional constraints provided by the $3 \lesssim z \lesssim 4$ sources effectively compensate for the large systematic uncertainties affecting each individual measurement, increasing the constraining power by $\sim 80\%$. 

As shown in Fig.~\ref{fig:wCDM-sys_muse_caminha}, these projected constraints are comparable to those reported by \citet{Caminha2022}, who combined five strong-lensing clusters with multiple source planes and considered statistical uncertainties only. By contrast, the constraints presented here explicitly include both statistical and systematic uncertainties. Despite this more conservative treatment, the Carousel Lens posterior remains competitive, being slightly narrower but exhibiting a higher degree of correlation. This correlation can be alleviated by combining our constraints with complementary probes, such as the CMB, which follow a different degeneracy direction and therefore enable significantly tighter joint constraints.

Figure~\ref{fig:wCDM-sys_sn1a_cmb_bao} illustrates that the Carousel Lens alone could provide constraints similar to those from the CMB, SNe~Ia, and BAO if high-resolution imaging of the newly identified high-redshift sources were available and comparable uncertainties could be achieved. In this regime, the Carousel Lens constraints remain nearly orthogonal to those from the CMB and BAO, highlighting strong complementarity, and similar in orientation to those from SNe~Ia, thus adding statistical power, while being subject to different systematics than either probe. In addition, these probes are affected by largely independent sources of systematic uncertainty, so their combination not only breaks degeneracies and increases statistical power but also enhances the overall robustness of the joint inference. This highlights the potential of multi-source-plane strong lensing, particularly in relaxed clusters, to provide competitive and independent cosmological constraints.

\begin{figure}[tbp]
    \centering
    \includegraphics[width=\linewidth]{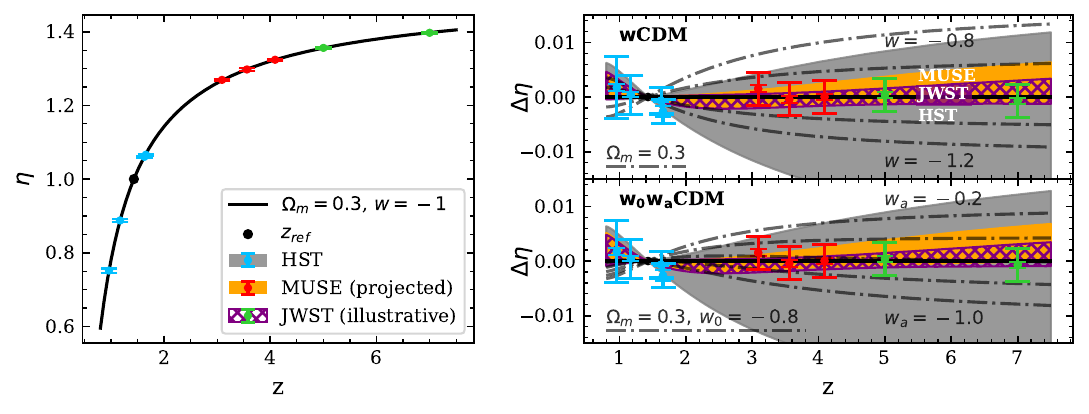}
    \caption{Strong-lensing Hubble diagram.
    \textit{Left}: Deflection ratio $\eta$ as a function of source redshift. The black line shows $\eta(z)$ for the concordance flat $\Lambda$CDM cosmology. The black point marks the reference redshift ($z = 1.432$, corresponding to sources 4 and 5). Cyan points denote the current HST constraints (sources 1, 3, 6, and 7), and red points correspond to MUSE-identified sources (sources 8, 11, and 12). 
    Green points indicate forecasted JWST sources at $z = 5$ and $z = 7$ for illustrative purposes (see text for details).
    Error bars indicate both $\sigma_{\rm stat}$ (smaller and thinner) and $\sigma_{\rm tot}$ (larger and thicker) for the HST sources, and the corresponding average uncertainties adopted for the MUSE and JWST sources. The $\eta$ values for the MUSE (and JWST forecasted) points are simulated assuming a flat $\Lambda$CDM cosmology with $\Omega_m = 0.3$, including a random scatter.
    \textit{Top right:} Difference between the measured deflection ratios and the concordance $\Lambda$CDM model prediction. Shaded regions show the $w$CDM constraints from Fig.~\ref{fig:wCDM-sys_hst_muse} projected into the $z$–$\eta$ plane, with the additional purple region indicating the forecast constraints including the $z=5$ and $z=7$ sources. Dash-dotted lines illustrate the variation of $\eta(z)$ as $w_0$ is varied from $-1.2$ to $-0.8$ in steps of $0.1$, with $\Omega_m$ fixed to 0.3.
    \textit{Bottom right}: Same as the top-right panel, but for the $w_0w_a$CDM parameterization. Dash-dotted lines show the evolution of $\eta(z)$ as $w_a$ varies from $-1.0$ to $-0.2$ in steps of $0.2$, assuming $\Omega_m = 0.3$ and $w_0 = -0.8$ (approximately the DESI DR2 centroid).
    }
    \label{fig:hubble-diagram}
\end{figure}

The source of the constraints can be visualized in an $\eta$–$z$ diagram, analogous to the SNe~Ia Hubble diagram, which we term the lensing Hubble diagram. In this plane, $\eta$ is obtained from strong-lens modeling and $z$ from spectroscopy. Each cosmological model traces a unique curve $\eta(z)$, all intersecting at $\eta(z_{\rm ref})=1$. Models are therefore indistinguishable near $z_{\rm ref}$ but diverge at higher or lower redshifts, where the constraining power is strongest. Fig.~\ref{fig:hubble-diagram} shows the lensing Hubble diagram with the current and projected constraints for the Carousel Lens for $w$CDM and $w_0w_a$CDM. High-redshift sources not yet included in the model provide much stronger constraints on $\eta(z)$, even with the same fractional uncertainties, while a larger number of sources reduces the scatter.

Fully exploiting the constraining power of the newly discovered $3 \lesssim z \lesssim 4$ MUSE sources would require multi-band constraints, a capability that is not currently implemented in our modeling framework and would significantly increase both model complexity and computational cost. We plan to incorporate this functionality in future work, as discussed in more detail in Subsection~\ref{sec:future}. In addition, achieving comparable uncertainties for the $3 \lesssim z \lesssim 4$ MUSE sources will require high-resolution follow-up observations, for example, with JWST/NIRCam. 

JWST observations are also expected to reveal strongly lensed sources at even higher redshifts, potentially up to $z \gtrsim 7$, and possibly within the current redshift gap at $2 \lesssim z \lesssim 3$. As illustrated in Fig.~\ref{fig:hubble-diagram}, the inclusion of two additional sources at $z = 5$ and $z = 7$ is expected to substantially tighten the constraints on $\eta(z)$ and, consequently, on cosmological parameters; this forecast is further discussed in Appendix~\ref{apx:additional-sources-forecast}. Since both SN~Ia and BAO measurements currently probe redshifts up to $z \lesssim 3$, extending the lensing Hubble diagram beyond this range would provide constraints on cosmology across a largely unexplored regime.

Finally, we acknowledge that this forecast assumes multi-lens-plane effects to be a subdominant source of uncertainty, as is generally the case for cluster-scale strong lenses. However, multi-plane lensing is expected to become increasingly important at higher source redshifts, with its specific impact depending on the relative alignment of the lens and source planes \citep[e.g.,][]{Wang2024}.
We will incorporate multi-plane lensing in our future modeling effort for the Carousel Lens as computational capabilities continue to improve (see Subsection~\ref{sec:future}).

\newpage
\subsection{Projected $w_0w_a$CDM constraints and comparison with other probes}

Evolving dark energy rises as an extension to the cosmological model that allows for a time evolution of the dark energy equation of state. While there is currently no robust evidence for dynamical dark energy, recent analyses of Type Ia supernovae from Pantheon+ \citep{Brout2022}, DES DR5 \citep{DES-DR5-SNe2024}, and Union3 \citep{Rubin2025}, as well as BAO measurements from DESI DR2 \citep{Abdul2025}, have reported a mild ($\sim 2$–$3\sigma$) tension with $\Lambda$CDM when combined with Planck 2018 CMB constraints, favoring $w_0 > -1$ and $w_a < 0$. However, this tension may be driven by residual systematics. For example, \citet{Popovic2025} show that recalibration of the DES DR5 sample reduces the tension from mild to weak. In contrast, \citet{Hoyt2026} find that recalibrating the Pantheon+ and Union3 samples leads to a more consistent level of tension across the three supernova analyses.

Independent constraints on $w_0\,\text{-}\,w_a$ are of high relevance in this context. The right bottom panel of the strong lensing Hubble diagram in Fig.~\ref{fig:hubble-diagram} shows how the Carousel Lens can constrain $w_0w_a$CDM cosmology through the deflection ratios of multiple sources. An evolving dark energy will produce a different evolution on $\eta$ compared to $w$CDM, which in the case of this lensing configuration is mostly noticeable at $z \lesssim 1$---currently weakly constrained by source 1---and $2 \lesssim z \lesssim 3$, where we currently lack sources. 

We expect the constraints from source~1 to improve with a more flexible mass parameterization and the inclusion of additional lensed images near the cluster center (in addition to 7d*, the inner images from source 11 and 13 from Paper~I, and the model-predicted counterimage for source 5), as its uncertainty is strongly driven by sensitivity to the inner density slope.

The absence of sources at $1.7 \lesssim z \lesssim 3$ likely reflects the wavelength coverage of MUSE, for which Ly$\alpha$ enters the observable range only blueward of $\sim 4800\,\text{\AA}$; this gap could be addressed with near-infrared spectroscopy or imaging, for example, with JWST.

\begin{figure}
    \centering
    \captionsetup[subfigure]{oneside,margin={0.8cm,0cm}}
    \subfloat[]{
        \includegraphics[height=0.31\linewidth]{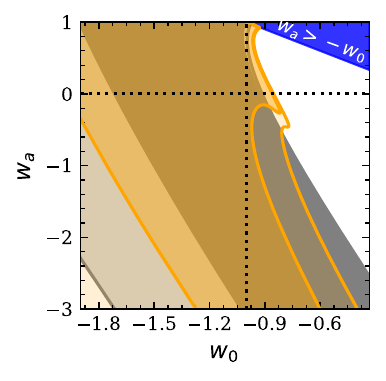}
        \label{fig:w0waCDM-hst-muse-jwst}
    }
    \captionsetup[subfigure]{oneside,margin={-5cm,0cm}}
    \subfloat[]{
        \includegraphics[height=0.31\linewidth]{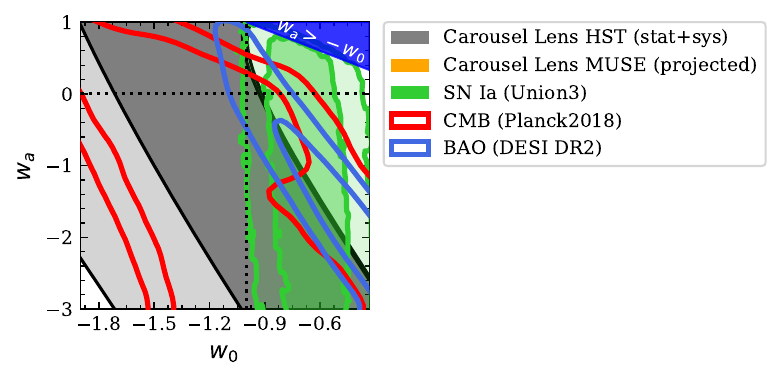}
        \label{fig:w0waCDM-sn-cmb-bao}
    }
    \caption{$w_0w_a$CDM constraints from the Carousel Lens. (a) Probability contours for $w_0$–$w_a$ constrained by the current HST data and predictions for MUSE sources. (b) Comparison with the $w_0$–$w_a$ constrains by the CMB \citep{planck2018}, SNe Ia \citep{Rubin2025}, and BAO \citep{Abdul2025}. Contours correspond to 68\% and 95\% confidence levels. The dashed line marks the concordance model ($\Omega_m=0.3$, $w_0=-1$, $w_a = 0$), and the blue region $w_a>-w_0$ is excluded as non-physical.}
\end{figure}

While higher-$z$ sources from MUSE could improve the constraints, the gain is not as large as in the $w$CDM case, as shown in Fig.~\ref{fig:w0waCDM-hst-muse-jwst}. This is because, at high redshift, the dependence of the deflection ratio $\eta$ on $w_0$ at fixed $w_a$ closely resembles its dependence on $w_a$ at fixed $w_0$ (see the comparison between the dot-dashed lines in the top- and bottom-right panels of Fig.~\ref{fig:hubble-diagram}), resulting in a strong degeneracy between the two parameters. As seen in Fig.~\ref{fig:w0waCDM-sn-cmb-bao}, the current constraints from the Carousel Lens are comparable to those from the CMB, which on its own does not provide stringent constraints but is typically combined with other probes such as SNe~Ia or BAO \citep{Brout2022, Rubin2025, Abdul2025}. These constraints can also be improved by considering a larger sample of multi-source-plane strong lenses \citep{Caminha2022, Sharma2023}.

Finally, given the possibility of evolving dark energy, the ability of the strong-lensing Hubble diagram to probe redshifts well beyond those currently accessible to SNe~Ia and BAO ($z \lesssim 3$), reaching out to $z \gtrsim 7$, can provide a uniquely powerful avenue for understanding the nature of dark energy.

\subsection{Future methodology improvements} \label{sec:future}

A number of avenues exist to reduce the dominant sources of systematic uncertainty in our analysis.  

\begin{itemize}
\item First, increasing the complexity of the mass model could mitigate the tension expected between inner and outer source images. 
Adopting more flexible parameterizations, such as NFW, generalized NFW, or broken power-law profiles, would allow for a radially varying slope. Although this may increase statistical uncertainty, the gain lies in reducing systematic biases. In particular, the five images (1a, 4e, 12e, 13e, 7d*) located very close to the cluster center (Fig.~\ref{fig:all-sources-jack})  should provide a strong constraint on the inner slope.  

\item Second, explicitly incorporating cluster members and substructures is physically motivated and improves the flexibility of the model. Modeling cluster galaxies at the pixel level is computationally prohibitive \citep[e.g.][]{Acebron2024}, but recent advances in efficient numerical approximations (Urcelay et al., in prep.) suggest that this cost can be significantly reduced, making the inclusion of larger numbers of galaxies feasible even in high-resolution surface brightness modeling.  

\item Third, the treatment of subhalo populations could be refined. Current subhalo scaling relations may be tightened by combining lensing constraints with kinematic measurements \citep[e.g.][]{Bergamini2019}, which is directly applicable to our MUSE and GMOS data. In addition, hierarchical modeling frameworks \citep[e.g.][]{Bergamini2021} can explicitly account for scatter in subhalo properties, converting part of the systematic uncertainty into a statistical one. This approach is expected to yield smaller overall uncertainties, as individual subhalos would be better constrained by kinematics than under the current random-scatter prescription.
For these reasons, we regard the current estimates presented in this work as the upper bound for systematic uncertainties.

\item Further improvements may be achieved through multi-band modeling, which enables the inclusion of additional background sources and tighter constraints. However, multi-band surface brightness modeling substantially increases computational cost, particularly GPU memory usage, and may require multi-GPU-node implementations, such as those under development for the \gigal pipeline (N.~Ratier-Werbin et al., in prep.).

\item For the highest-redshift sources, a multi-lens-plane formalism may become necessary, though its relevance as a dominant source of systematics remains to be established.  

\item Finally, further optimization could be achieved by adopting a self-consistent cosmological framework within the lens model, rather than a two-step inference (first deriving $\eta$ parameters and subsequently cosmological parameters). Controlled tests with simulations would also help to quantify both systematics and statistical robustness. 
\end{itemize}
In summary, we anticipate a significant tightening of the cosmological constraints through the improvement avenues discussed in this subsection.

\section{Conclusions} \label{sec:conclusion}
In this work, we present an updated strong-lensing model of the Carousel Lens, incorporating two newly identified multiply imaged sources and two additional mass components. The resulting model successfully reproduces the observed image positions, morphologies, and surface-brightness features, with the exception of the radial arc associated with source 7. Moreover, the model predicts the redshifts of sources 6 and 7 in good agreement with available spectroscopic measurements, providing an additional validation of the reconstruction.

Using this lens model, we derive cosmological constraints from deflection-ratio measurements, probing both the dark matter density and the equation of state of dark energy. The seven background sources detected in the HST imaging, spanning the redshift range $0.96 \lesssim z \lesssim 1.6$, yield four deflection-ratio constraints. We account for both statistical uncertainties and systematic effects through dedicated simulations. Within a $w$CDM framework, we obtain constraints consistent with $\Lambda$CDM. Our results, including systematic uncertainties---which dominate over the statistical uncertainties by a factor of five---are comparable to those derived from individual cluster lenses in the literature that account for statistical uncertainties alone. Specifically, we find $\Omega_m = 0.34^{+0.16}_{-0.13}$ and $w = -1.31^{+0.35}_{-0.32}$, with a strong non-linear correlation between the two parameters. Combining these constraints with CMB data breaks this degeneracy, yielding $\Omega_m = 0.26 \pm 0.03$ and $w = -1.2 \pm 0.1$.

Recently discovered multiply imaged systems identified by Paper~I extend the available source redshift range to $z \sim 4$. We show that, if systematic uncertainties for these higher-redshift sources remain comparable to those inferred from the HST sample, the constraining power of the Carousel Lens on $w$CDM cosmology could improve by approximately $80\%$. Under these assumptions, a single well-modeled cluster lens could yield uncertainties comparable to those from established cosmological probes such as the CMB, Type Ia supernovae, and BAO. Importantly, the Carousel Lens constraints are complementary to these probes in two key respects: first, they exhibit different (often nearly orthogonal, e.g., relative to the CMB) parameter degeneracies in the $\Omega_m$–$w$ plane, enhancing joint statistical power (even when the orientations are similar, e.g., relative to SNe~Ia, the Carousel Lens constraints nearly doubles the constraining power); and second, they are affected by largely independent sources of systematic uncertainty. 

For evolving dark energy models, the Carousel Lens provides constraints on the $w_0w_a$CDM parameters that are comparable in precision to those from the CMB alone. While these constraints are weaker than those obtained from SNe~Ia and BAO---owing to the strong degeneracy between $w_0$ and $w_a$ inherent to deflection-ratio measurements---they nonetheless offer valuable complementary information, statistical power, and different systematics for joint analyses. Furthermore, if systematic uncertainties can be substantially reduced with future modeling efforts---as appears feasible---the constraints from the Carousel Lens could improve markedly (even for $w_a$) and rival those from BAO.

This work represents the first cosmological constraint derived from distance ratios measured in a galaxy cluster lens modeled using extended surface-brightness information, whereas previous studies, such as \citet{Caminha2022}, relied on position-based modeling. At present, our results are dominated by systematic uncertainties, which we estimate through simulations. Reducing these systematics will require a more flexible mass-model parameterization, as well as the incorporation of additional constraints such as cluster-member kinematics and multi-wavelength imaging.

In particular, JWST imaging of the Carousel Lens would enable the inclusion of the already discovered $3 \lesssim z \lesssim 4$ MUSE sources in a surface-brightness model through multi-band pixel-level constraints. Beyond these known systems, JWST may also reveal even higher-redshift lensed sources (e.g., $z \gtrsim 5$), further extending the redshift leverage of the lensing ``Hubble diagram'' and increasing sensitivity to the dark energy equation of state. Realizing this potential will require increased---but still tractable---computational resources to accommodate the larger data volume and explore more flexible lens models.

More broadly, the observational and computational requirements for this approach are modest when compared to those needed for large spectroscopic and photometric surveys targeting thousands of supernovae or mapping large-scale structure over cosmological volumes with millions of galaxies. While strong-lensing cosmography relies on detailed modeling of individual systems, targeted follow-up observations and careful modeling enable systematic uncertainties to be identified, quantified, and progressively reduced. As demonstrated here, this makes cluster strong lensing a promising and cost-effective complementary and competitive probe for precision cosmology.

\section*{Acknowledgments} \label{sec:acknowledgments}
This work was supported in part by the Director, Office of Science, Office of High Energy Physics of the US
Department of Energy under contract No. DE-AC025CH11231. This research used resources of the National Energy
Research Scientific Computing Center (NERSC), a U.S. Department of Energy Office of Science User Facility operated
under the same contract as above and the Computational HEP program in The Department of Energy’s Science Office
of High Energy Physics provided resources through the “Cosmology Data Repository” project (Grant \#KA2401022).  This research was supported by the Australian Research Council Centre of Excellence for All Sky Astrophysics in 3 Dimensions (ASTRO 3D), through project number CE170100013.  
The work of A.C. is supported by NOIRLab, which is managed by the Association of Universities for Research in Astronomy (AURA) under a cooperative agreement with the National Science Foundation. 
X.H. acknowledges the University of San Francisco Faculty Development Fund.
T.J. and K.V.G.C. gratefully acknowledge financial support from the National Science Foundation through grant AST-2108515, NASA through grant HST-GO-16773, the Gordon and Betty Moore Foundation through Grant GBMF8549, and from a Dean’s Faculty Fellowship.  We thank Ned Taylor and Michelle Cluver of the Swinburne University of Technology for providing reduced VISTA 4MOST Hemisphere Survey data.

\software{Lenstronomy \citep{birrer2018, lenstronomyII},
          NumPy \citep{numpy}, 
          SciPy \citep{scipy},
          Matplotlib \citep{matplotlib}, 
          Astropy \citep{astropy2013, astropy2018, astropy2022}, 
          Jupyter \citep{kluyver2016},
          TensorFlow \citep{TensorFlow2015,TensorFlowProb2017},
          JAX \citep{jax2018github},
          Photutils \citep{photutils}.
          }

\bibliography{sample631}{}
\bibliographystyle{aasjournal}

\appendix

\section{MCMC convergence}
\label{apx:chains}
\begin{figure}[tbhp]
    \centering
    \includegraphics[width=0.96\linewidth]{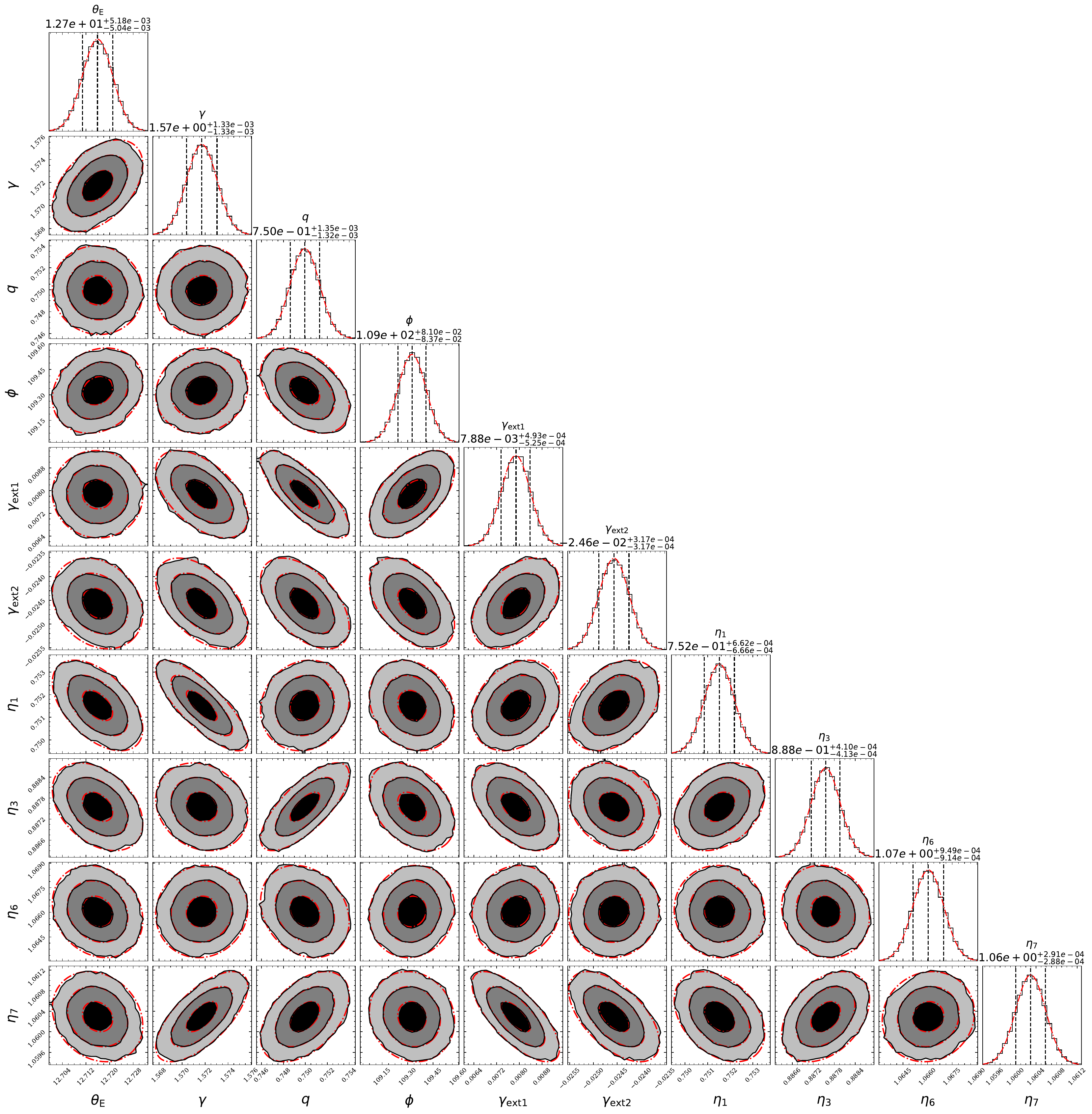}
    \caption{Lens model posterior distribution. 
    Corner plot showing the posterior distributions of the lens-model parameters sampled with the MCMC, marginalizing over the source positions. The parameters shown are the main halo Einstein radius $\theta_E$ (arcseconds), logarithmic slope $\gamma$, axis ratio $q$, position angle $\phi$ (degrees east of north), the two components of the external shear $\gamma_{\rm ext,1}$ and $\gamma_{\rm ext,2}$, and the deflection ratios for sources 1, 3, 6, and 7 ($\eta_i$). Gray contours correspond to the 1, 2, and 3$\sigma$ confidence levels. For each parameter, the marginalized one-dimensional distribution is shown along the diagonal, with vertical dashed lines and labels indicating the mean value and its $1\sigma$ uncertainty. Red dot-dashed contours and histograms correspond to a multivariate normal distribution with mean and covariance matched to the MCMC samples, illustrating that the 2-D marginal posteriors are well approximated by Gaussians.}
    \label{fig:mcmc_cornerplot}
\end{figure}

\begin{figure}[tbhp]
    \centering
    \includegraphics[width=\linewidth]{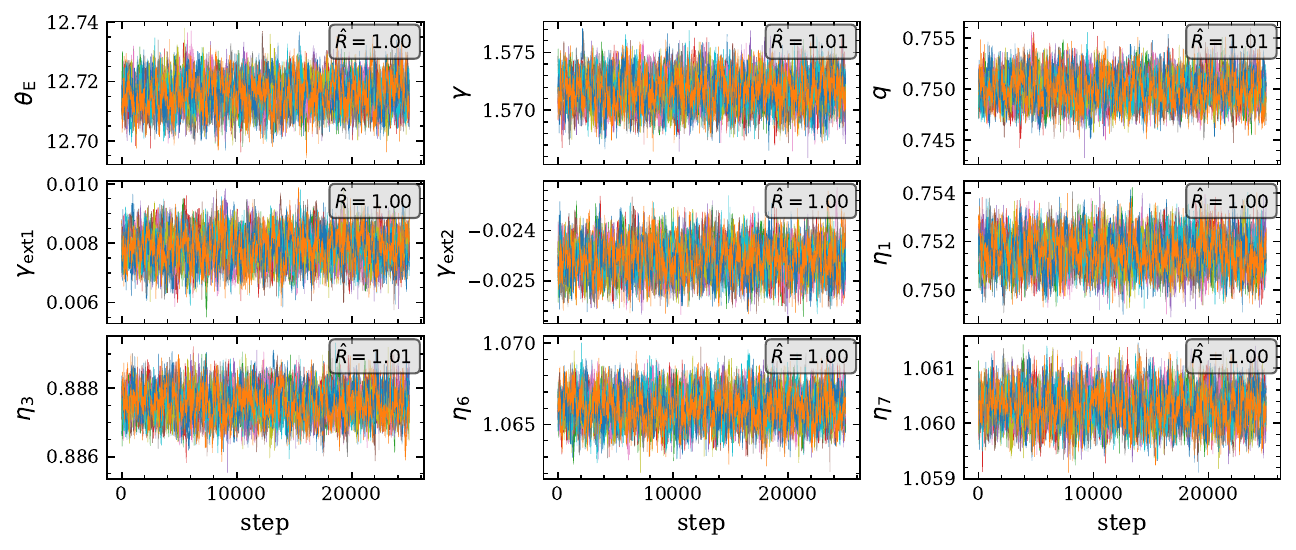}
    \caption{MCMC trace plots.
    Each panel shows the MCMC trace for one model parameter. The lines represent the sampled parameter values over the 30,000 MCMC steps, excluding the initial 5\,000-step burn-in phase. Different colors correspond to the 12 independent chains. The $\hat{R}$ convergence diagnostic is shown at the top-right corner of each panel. The traces show good mixing, low autocorrelation, and stable sampling behavior across all parameters.}
    \label{fig:mcmc_traceplot}
\end{figure}

\begin{figure}[tbhp]
    \centering
    \includegraphics[width=0.6\linewidth]{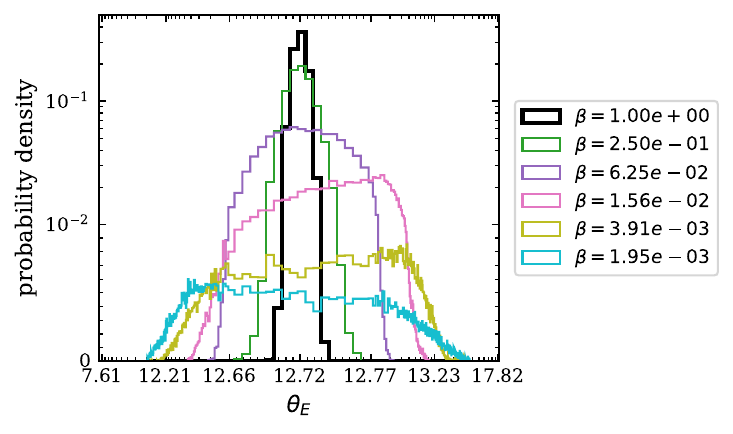}
    \caption{Tempered posterior distributions.
    The contour plot shows the marginalized distribution of the main halo $\theta_E$ parameter at different inverse temperature levels ($\beta$). Black contours correspond to the posterior distribution ($\beta = 1$), while lower values of $\beta$ indicate progressively ``hotter'' annealed chains that explore a wider region of parameter space. Note that both axes are log-spaced for visualization purposes.}
    \label{fig:mcmc_tempering}
\end{figure}

The lens-model parameter inference is performed using MCMC sampling implemented with the TensorFlow Probability \texttt{Python} package \citep{TensorFlow2015, TensorFlowProb2017}. To efficiently explore the potentially complex posterior landscape, we adopt a parallel tempering scheme \citep{Kofke2002, Earl2005}, in which multiple MCMC chains are run simultaneously at different annealing (inverse temperature) levels.

In this framework, the likelihood is weighted by an inverse temperature parameter $\beta \leq 1$. Chains with $\beta = 1$ sample the target posterior distribution, while chains with $\beta < 1$ sample progressively flatter versions of the likelihood, allowing them to explore a wider region of parameter space. Transitions within each chain are performed using an HMC kernel, preconditioned with the covariance matrix inferred from SVI. In addition, chains at neighboring temperature levels are allowed to swap states, propagating the information from ``hotter'' states to the posterior.

We employ a grid of MCMC chains with ten inverse-temperature levels, geometrically spaced as $\beta_i = 0.5^i$ for $i \in {0,\ldots,9}$. At each temperature level, we run 12 independent chains, resulting in a total of 120. The sampler is run for 30,000 steps, and the first 5000 steps of each chain are discarded as burn-in. The final posterior is constructed from the remaining 25,000$\, \times 12$ samples drawn from the chains at $\beta = 1$.

The marginalized posterior distributions of the lens-model parameters are shown in Fig.~\ref{fig:mcmc_cornerplot}, while the corresponding MCMC trace plots are presented in Fig.~\ref{fig:mcmc_traceplot}. The traces exhibit stable behavior over the sampling period and show no evidence of long-term drifts or poor mixing. We further assess convergence using the potential scale reduction factor $\hat{R}$ \citep{Gelman1992}, finding values well below the recommended threshold $\hat{R}<1.1$ \citep{Gelman2014}, indicating robust convergence and consistent sampling across chains.

To illustrate the role of parallel tempering in exploring the parameter space, Fig.~\ref{fig:mcmc_tempering} shows the marginalized distribution of the main halo $\theta_E$ for several inverse-temperature levels. While the posterior distribution ($\beta = 1$) is narrowly peaked around $\theta_E \simeq 12.7''$, the hotter chains explore a substantially broader range, spanning approximately $11.5''$ to $14.5''$. Our posterior distribution remains unimodal within this wide region, with no evidence for additional modes within the parameter space explored by the tempered chains.

\section{Systematic Uncertainties}
\label{apx:simulations}
\subsection{Subhalos and Density profile}
\label{apx:subhalo-scatter}

We use dedicated simulations to quantify the impact of assumptions about the main cluster-scale mass profile and the treatment of subhalos associated with cluster members. In total, we generate 11 simulated realizations of the Carousel Lens, all adopting the same source surface-brightness models as in our fiducial reconstruction but varying the underlying mass distribution. All simulations assume a flat $\Lambda$CDM cosmology with $\Omega_m = 0.3$.

Cluster members are selected using the red sequence of early-type galaxies. We fit a linear relation to the $F200LP - F140W$ color–magnitude diagram with $3\sigma$ clipping, and classify as members the 95 galaxies brighter than $F140W = 22.5,\mathrm{mag}$ and within $0.5,\mathrm{mag}$ of the sequence. All spectroscopically confirmed members from Paper~I fall within this selection. While the resulting sample is likely contaminated by non-members, this is expected to increase the inferred systematic uncertainties rather than bias the results.

\begin{figure}[tbhp]
    \centering
    \includegraphics[width=0.8\linewidth]{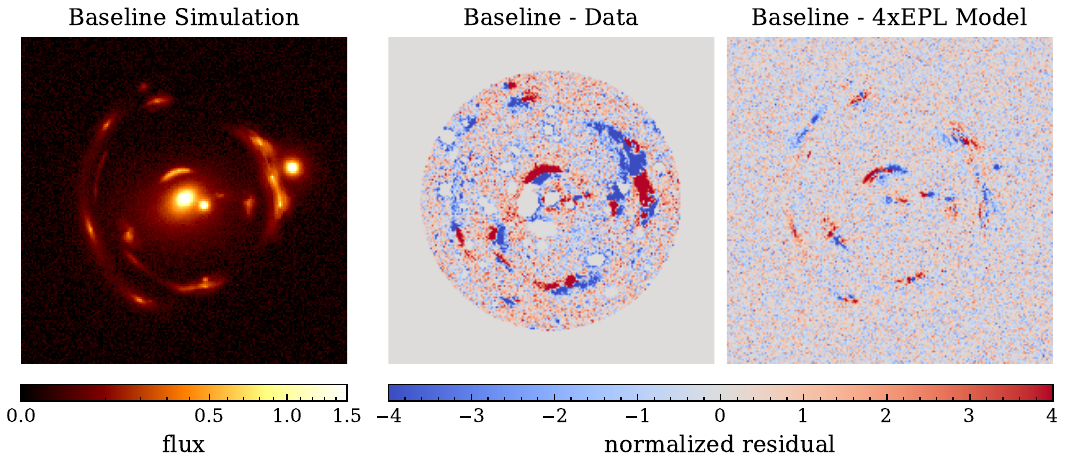}
    \caption{Baseline NFW+subhalos simulation.
    \textit{Left}: Baseline simulated surface brightness of the cluster core, assuming an NFW profile for the main halo and dPIE subhalos following the scaling relations of \citet{Bergamini2019} without intrinsic scatter, except for members $L_d$ and $L_e$, which are modeled using the same mass profiles as in our fiducial lens model. \textit{Middle}: Difference between the observed HST F140W image and the baseline simulation. \textit{Right}: Difference between the baseline simulation and the best-fit 4×EPL model, illustrating the impact on the surface brightness of adopting a different mass-model parameterization.}
    \label{fig:syserror_base_model}
\end{figure}

The first simulation realization, hereafter referred to as the baseline simulation, replaces the EPL profile of the main cluster halo with an NFW profile. The scale radius is fixed to $R_s = 70''$, consistent with the mass–concentration–redshift relation of \citet{Diemer2019} for a halo of mass $M_{200c} = 10^{15}\,M_\odot$ at $z = 0.49$. With this choice of $R_s$, the logarithmic slope of the profile varies from $-1$ in the inner regions to $\sim -2$ within $30''$, broadly consistent with the effective slope $\gamma \sim 1.6$ inferred in our fiducial model. The characteristic density $\rho_0$ is set such that the Einstein radius of source 4 matches the observed value. In addition, each color-selected cluster member is assigned a subhalo described by a dPIE profile following the scaling relations of \citet{Bergamini2019}, with no intrinsic scatter. This baseline simulation is then fitted with our fiducial four-EPL (4×EPL) model, which serves as a reference point for assessing the impact of subhalo scatter. A comparison between the baseline simulation, the data, and the fitted model is shown in Fig.~\ref{fig:syserror_base_model}.

\begin{figure}[tbhp]
    \centering
    \includegraphics[width=0.45\linewidth]{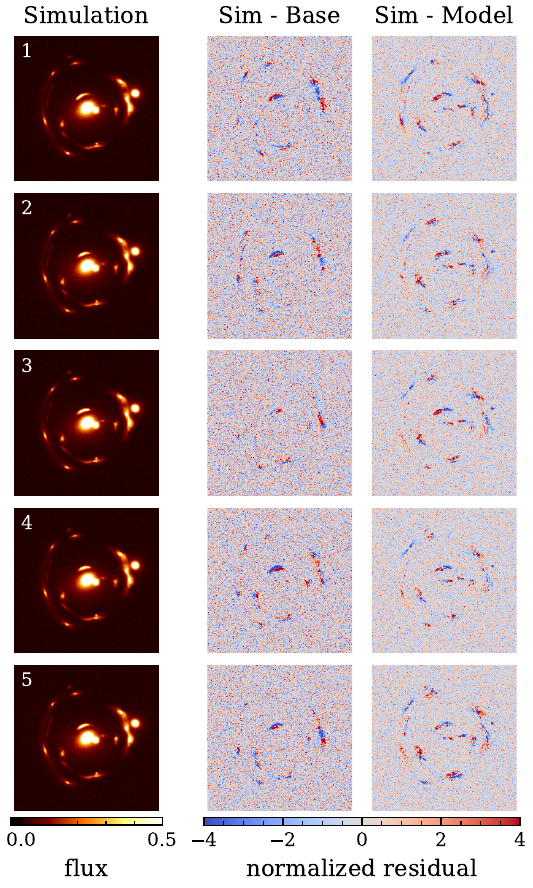}
    \hspace{0.5cm}
    \includegraphics[width=0.45\linewidth]{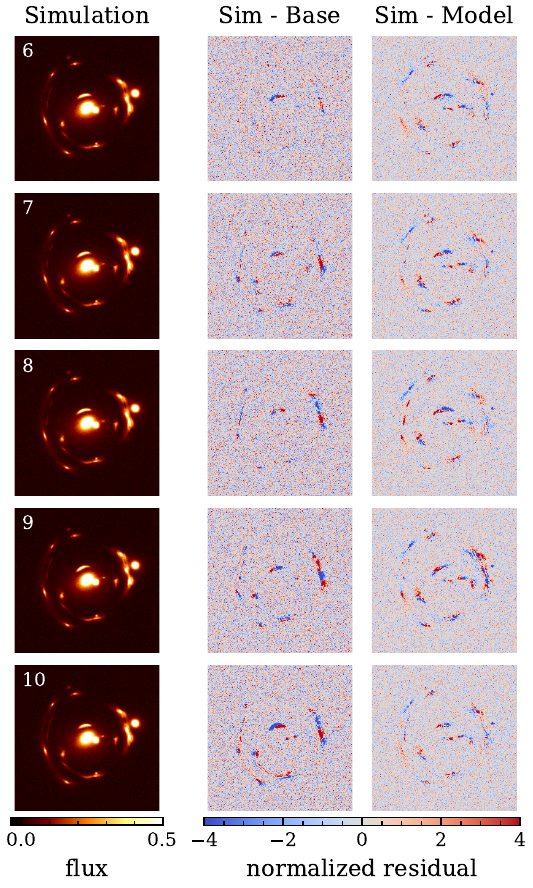}
    \caption{Impact of subhalo mass scatter.
        Illustration of the effect of intrinsic scatter in subhalo mass scaling relations on the lens model, based on ten simulated realizations with a 15\% scatter. The figure is divided into two halves, showing simulations 1–5 (left) and 6–10 (right). Within each half, the left column displays the simulated lensed surface-brightness distribution, the middle column shows the difference between the simulation with scatter and the baseline (no-scatter) simulation, and the right column shows the difference between the scattered simulation and the best-fit 4×EPL model, highlighting the impact of subhalo scatter on the recovered lens model. The intrinsic scatter does not alter the overall image configuration but instead produces small shifts in image positions and morphologies.}
    \label{fig:syserror_scatter}
\end{figure}

\begin{figure}[tbhp]
    \centering
    \includegraphics[width=0.5\linewidth]{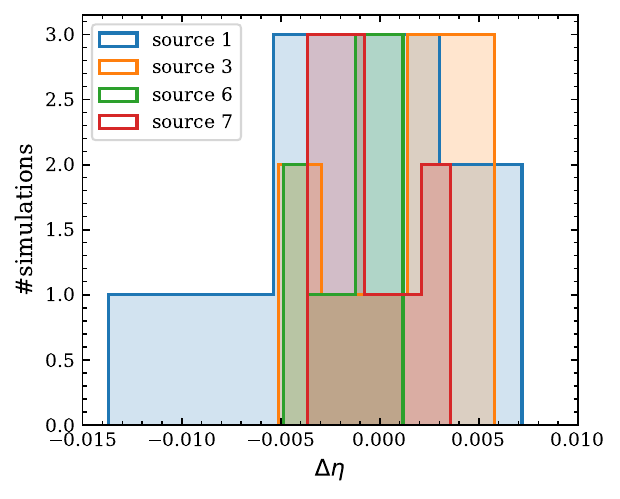}
    \caption{Distribution of the systematic error in the deflection ratio $\Delta\eta$ for each source over the ten simulations.}
    \label{fig:syserror_scatter_hist}
\end{figure}

Starting from this baseline, we generate ten additional realizations by introducing a 15\% random scatter in the subhalo scaling relations of \citet{Bergamini2019}. Each simulated system is fitted with the 4×EPL model, initialized from the best-fit baseline solution and allowing the deflection ratios of the sources to vary. The simulated images and corresponding model residuals are shown in Fig.~\ref{fig:syserror_scatter}. We then estimate the systematic uncertainty associated with the assumed main-halo profile and the mass and scatter of the subhalos by comparing the deflection ratios inferred from the data with those recovered from the simulations. Specifically, for each source $i$ and each realization $j$, we compute the deviation in the deflection ratio as $\Delta\eta_{i,j} = \eta_{{\rm model},i,j} - \eta_{\rm fiducial,i}$, where $\eta_{{\rm model},i,j}$ is obtained from the model fitted to the $j$-th simulation, and $\eta_{\rm fiducial,i}$ corresponds to the value implied by the cosmology assumed in the simulation at redshift $z_i$. The systematic uncertainty $\sigma_{\rm sys,i}$ for each source is then defined as the root-mean-square (${\rm RMS}_i$) of $\Delta \eta_{i,j}$ across the ten realizations. The resulting distributions of $\Delta\eta_{i,j}$ are shown in Fig.~\ref{fig:syserror_scatter_hist}.

As shown in Fig.~\ref{fig:syserror_base_model}, the differences between the 4×EPL model and the baseline simulation are generally comparable in amplitude to the model–data residuals (Fig.~\ref{fig:model-residual}), with the notable exception of source 1, which exhibits significantly larger deviations. This behavior persists in the realizations that include random subhalo scatter (Fig.~\ref{fig:syserror_scatter}) and can be traced to the flatter inner density slope of the NFW profile ($\rho \propto r^{-1}$) relative to the approximately constant $\sim 1.6$ logarithmic slope enforced by the EPL parameterization. As a consequence, the systematic uncertainty associated with source 1 is roughly a factor of two larger than that of the other sources. For the remaining systems, the introduction of scatter into the subhalo scaling relations typically produces residuals comparable to or larger than those of the fiducial model applied to the data. These simulations therefore indicate that intrinsic scatter in the subhalo mass–luminosity relations is the dominant source of uncertainty for most sources, with the notable exceptions of source 1 (and likely the radial image of source 7), where sensitivity to the inner mass-profile slope becomes important.

\subsection{Multi-lens plane}
\label{apx:multiplane}
To estimate the systematic impact of multi-plane lensing induced by the mass of background sources at $z > 0.49$, we perform a set of simulations in which additional lens planes are introduced into our fiducial lens model. The masses of these secondary deflectors are estimated from the kinematic properties of the lensed sources.

Since multi-lens-plane is currently under development for the \gigal{} pipeline, we carry out this analysis using the \texttt{Lenstronomy} package \citep{birrer2018, Birrer2021}. We employ its multi-plane lensing formalism, adopting a reference redshift $z_{\rm ref}^{\rm multi} = 5$. The parameters of our fiducial single-plane model---originally defined with a reference redshift $z_{\rm ref} = 1.432$---are transformed to this new reference by rescaling them according to the deflection ratio $\eta_{\rm multi}$ between the two reference planes. In particular, the Einstein radius of each EPL component is scaled by $\eta_{\rm multi}^{1/(\gamma-1)}$, while the external shear amplitude is scaled linearly with $\eta_{\rm multi}$. All other lens parameters remain unchanged.

The resulting multi-plane model includes three lens planes. The primary lens plane at $z = 0.49$ has the same mass distribution as our fiducial single-plane model and represents the cluster-scale deflector. Two additional lens planes are placed at $z = 1.166$ and $z = 1.432$, corresponding to sources 3 and 4, respectively. Each of these planes is modeled as a singular isothermal ellipsoid (SIE), with position and ellipticity matched to the reconstructed source-plane surface-brightness morphology.

Other background sources are not included in the multi-plane modeling, as their expected lensing impact is subdominant compared to that of sources 3 and 4. Sources 1 and 5 show no significant rotational or dispersion signal in the MUSE data and are fainter and more compact in the source-plane reconstruction. Source 6 lies at a redshift very close to that of source 7, and therefore has a negligible lensing effect regardless of its mass. While sources 6 and 7 may influence higher-redshift systems detected in the MUSE data, the lack of strong emission lines and the presence of only weak absorption features prevent a reliable kinematic mass estimate; their contribution will be explored in future work.

\begin{figure}[tbhp]
    \centering
    \includegraphics[height=0.40\linewidth]{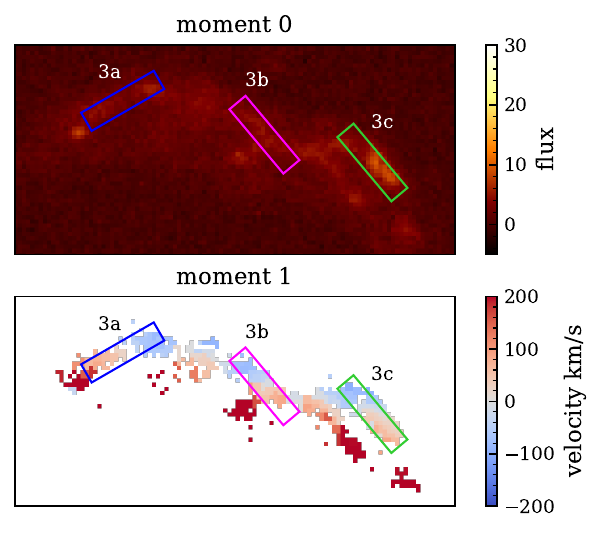}
    \includegraphics[height=0.40\linewidth]{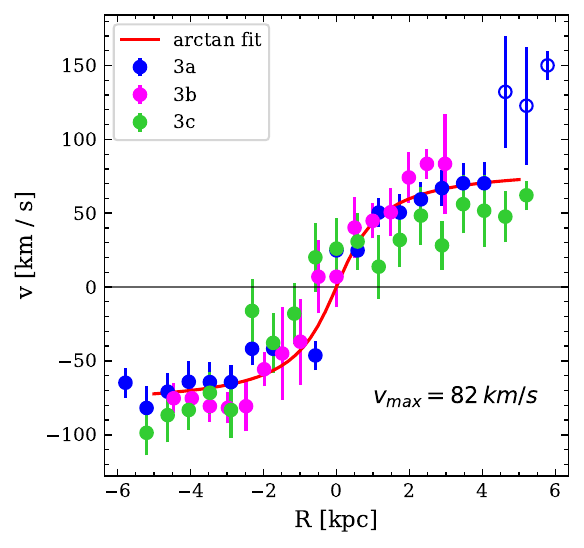}
    \caption{Kinematics of source 3.
    \textit{Left}: Flux and line-of-sight velocity maps of source 3 traced by [O II] emission at $z = 1.166$ from the MUSE observations. Colored rectangles indicate the slits used to extract the rotation curves for each lensed image. \textit{Right}: Rotation curve obtained by integrating the velocity field along each slit. The red curve shows the best-fit arc-tangent model. A subset of data points exhibiting large velocity offsets ($\sim100$–$300\ \mathrm{km,s^{-1}}$), likely associated with an outflow or a companion galaxy, are shown as open circles and excluded from the fit.}
    \label{fig:syserror_kinematics_s3}
\end{figure}

\begin{figure}[tbhp]
    \centering
    \includegraphics[height=0.40\linewidth]{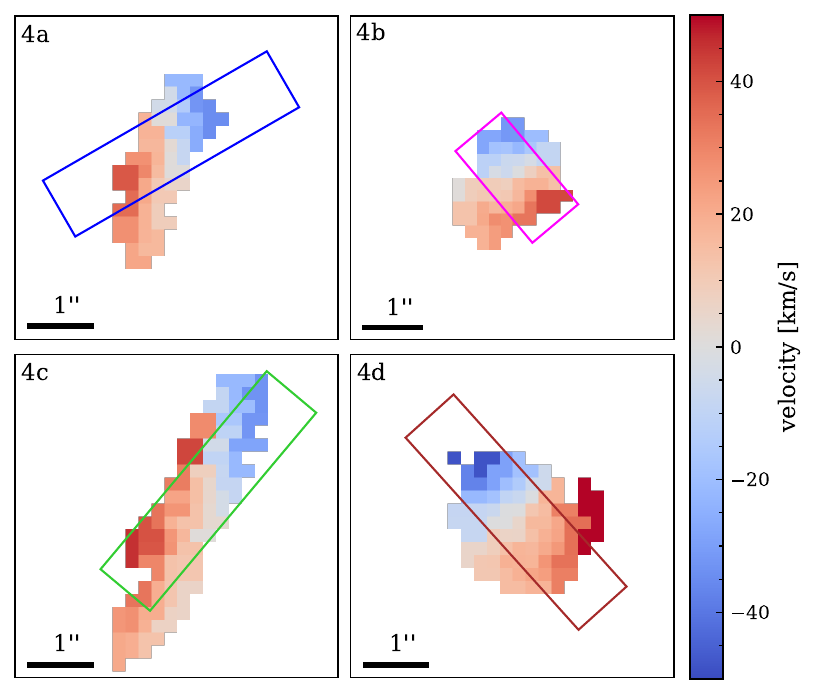}
    \includegraphics[height=0.40\linewidth]{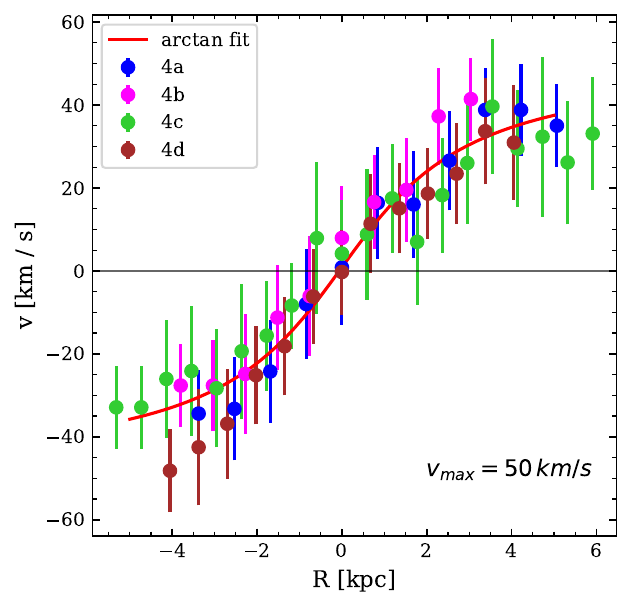}
    \caption{Kinematics of source 4.
    \textit{Left}: Line-of-sight velocity maps of source 4 images traced by [O II] emission at $z = 1.432$ from the MUSE observations. Colored rectangles indicate the slits used to extract the rotation curves for each lensed image. \textit{Right}: Rotation curve obtained by integrating the velocity field along each slit. The red curve shows the best-fit arc-tangent model. }
    \label{fig:syserror_kinematics_s4}
\end{figure}

We estimate the masses of sources 3 and 4 using their rotation signals detected in the deep MUSE observations. For each source, we extract a rotation curve from the image-plane kinematic map by integrating the velocity field over a $1''$-wide slit centered on the source and aligned with its kinematic major axis. Distances from the kinematic center are mapped from the image plane to the source plane using the lens model. An arc-tangent function is then fitted to the resulting rotation curve to infer the maximum rotation velocity, $v_{\rm max}$. The kinematic maps and rotation courves for sources 3 and 4 are shown at Figs. \ref{fig:syserror_kinematics_s3} and \ref{fig:syserror_kinematics_s4} respectively.

Converting the rotation velocity into a mass requires an assumption about the inclination angle $i$, which is the dominant source of uncertainty, particularly given the sensitivity of the inferred source ellipticity to lensing shear. To bracket this uncertainty, we perform the analysis assuming three representative inclination angles: $i = 30^\circ$, $50^\circ$, and $70^\circ$. The enclosed mass at radius $R = 3,{\rm kpc}$ in the source plane is then computed as
\begin{equation}
    M(R) = \frac{v(R)^2\,R}{G\,\sin(i)}
\end{equation}
where $G$ is the gravitational constant and $v(R)$ is the rotation velocity at radius $R$.

\begin{figure}[tbhp]
    \centering
    \includegraphics[width=0.8\linewidth]{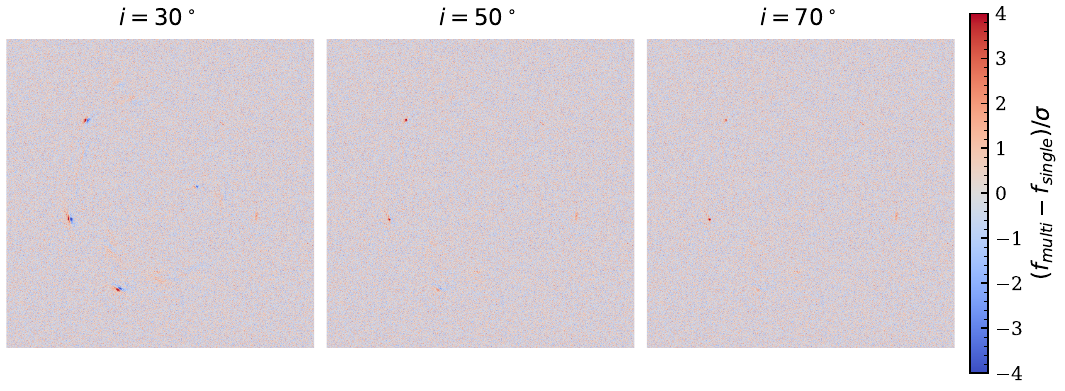}
    \caption{Multi–lens-plane effects.
            Difference between a single–lens plane model and a multi–lens plane simulation of the Carousel Lens. The single–lens plane model corresponds to the fiducial mass model used throughout this work, with all mass placed at the main lens redshift $z = 0.49$. The multi–lens plane model includes the same mass distribution at the main lens plane, and additionally incorporates SIE mass components at the redshifts of sources 3 and 4. The positions and geometric parameters of these additional deflectors follow the source-plane surface-brightness models, while their masses are estimated from kinematic measurements. The three panels illustrate the impact of different inclination-angle assumptions---$30^\circ$, $50^\circ$, and $70^\circ$---on the inferred mass of sources 3 and 4, highlighting the resulting variation in the multi–lens-plane lensing signal.}
    \label{fig:syserror_multiplane}
\end{figure}

Finally, for each source we compute the Einstein radius of an SIE lens placed at the corresponding redshift $z_i$ and referenced to $z_{\rm ref}^{\rm multi}$, such that it encloses the same mass within radius $R$:
\begin{equation}
    \theta_{E,i} = \frac{4\,G\,M(R)}{c^2\,R}
    \frac{D_{ls}(z_i, z_{\rm ref}^{\rm multi})}{D_s(z_{\rm ref}^{\rm multi})}
\end{equation}
The differences between the resulting multi–lens-plane models---computed for each assumed inclination---and our fiducial single–lens-plane model are shown in Fig.~\ref{fig:syserror_multiplane}. Overall, the impact of multi-plane lensing is found to be negligible for most sources, with the exception of source 7. Notably, source 6, which lies at a similar redshift but is located much closer to sources 3 and 4 in the source plane (see Fig.~\ref{fig:model-residual}), exhibits significantly smaller residuals than source 7. We therefore attribute the larger residuals observed for source 7 to its bright and compact surface-brightness profile, which makes it particularly sensitive to small positional offsets. Even in this case, the effect is only appreciable for the lowest inclination-angle assumption ($i = 30^\circ$) and remains small compared to the residuals induced by intrinsic scatter in the subhalo population (Appendix~\ref{apx:subhalo-scatter}). We thus conclude that multi–lens-plane effects constitute a subdominant source of systematic uncertainty for the sources considered in the HST data.

\section{Image configuration and cosmological constraints}
\label{apx:geometry}
\begin{figure}
    \centering
    \subfloat[quad]{
        \includegraphics[width=0.325\linewidth]{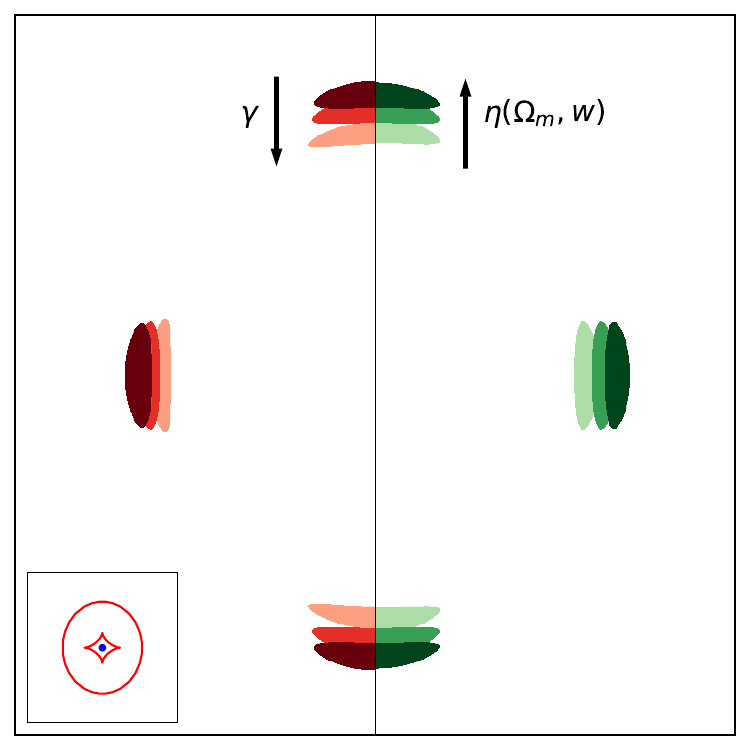}
        \label{fig:geometry-cosmo-quad}%
    }
    \subfloat[double]{%
        \includegraphics[width=0.325\linewidth]{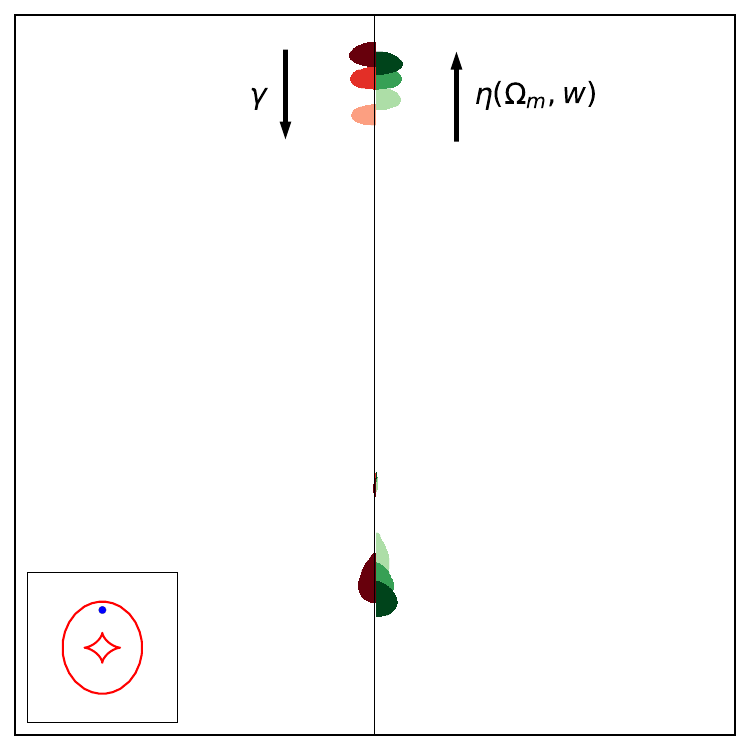}
        \label{fig:geometry-cosmo-double}%
    }
    \subfloat[cusp]{%
        \includegraphics[width=0.325\linewidth]{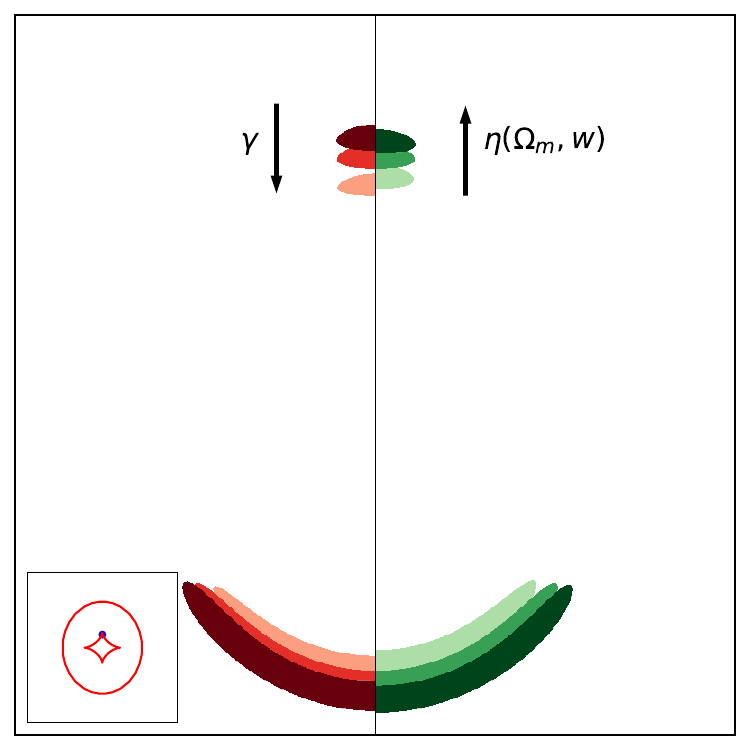}
        \label{fig:geometry-cosmo-cusp}%
    } \\
    \subfloat[cusp with radial image]{
        \includegraphics[width=0.325\linewidth]{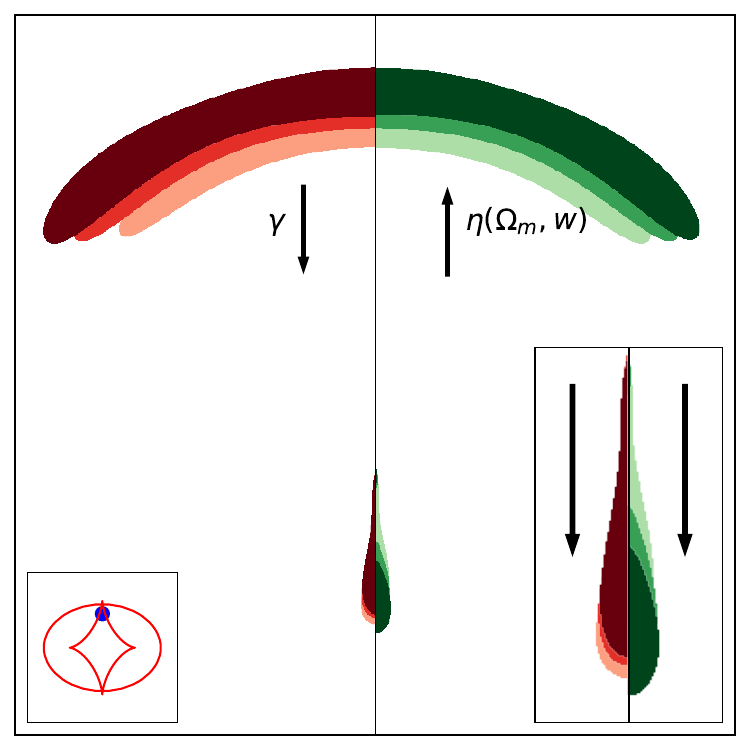}
        \label{fig:geometry-cosmo-radial}%
    }
    \subfloat[fold (slope)]{%
        \includegraphics[width=0.325\linewidth]{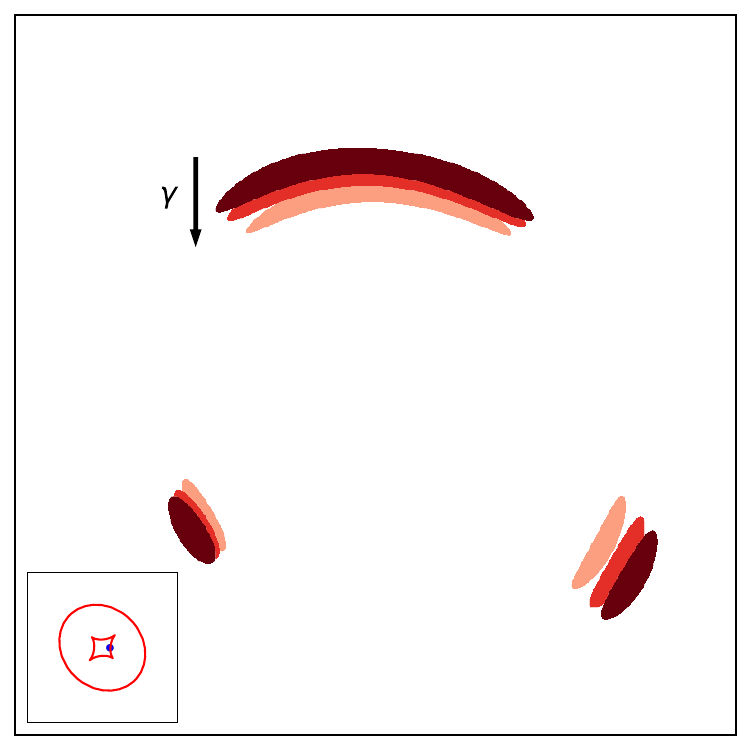}
        \label{fig:geometry-cosmo-fold-slope}%
    }
    \subfloat[fold (cosmology)]{%
        \includegraphics[width=0.325\linewidth]{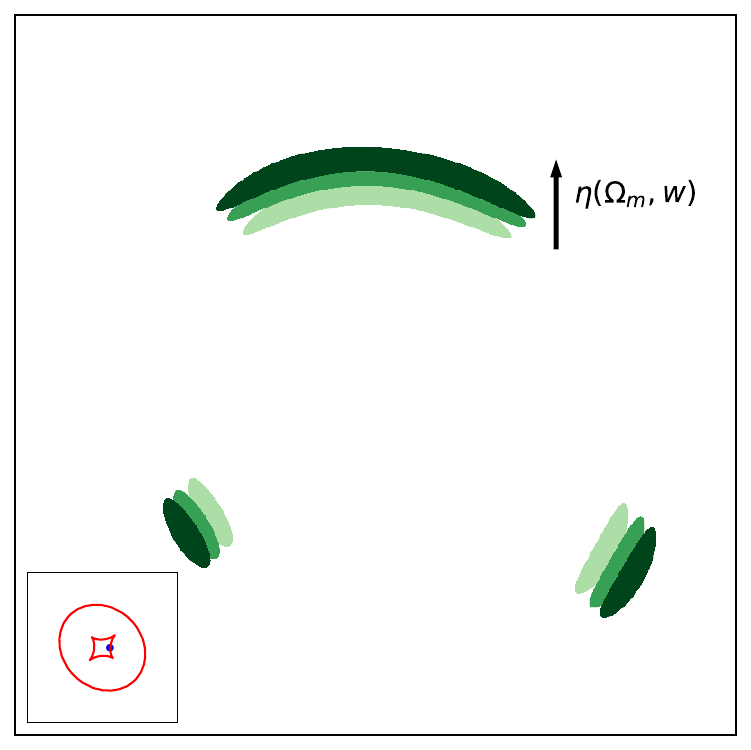}
        \label{fig:geometry-cosmo-fold-cosmo}%
    }
    \caption{%
        Degeneracy between density slope and cosmology in multi-source plane strong lensing.
        Qualitative illustration of the correlation between the power-law density slope $\gamma$ and the deflection ratio $\eta(\Omega_m, w)$ in a $w$CDM cosmology for different multiple-image configurations. Each panel shows a distinct lensing configuration for a circular source with a stepwise surface-brightness profile, with images generated by varying either the mass-profile slope $\gamma$ or the deflection ratio $\eta$. Insets in the lower-left corners display the corresponding critical curves and caustics, with the source indicated by a blue circle.
        Red regions show the lensed images obtained by varying $\gamma$ over the values 1.65, 1.7, and 1.8 (lighter shades correspond to larger $\gamma$), while green regions show images produced by varying $\eta$ through changes in cosmology, with $(\Omega_m, w) = (0.1, -1.5)$, $(0.3, -1)$, and $(0.5, -1.65)$ in order of increasing $\eta$ (darker shades indicate larger $\eta$). These parameter values are chosen for illustrative purposes only.
        Each panel is split along the axis of symmetry to contrast the effects of varying $\gamma$ and $\eta$, except in panels (e) and (f), which show fold configurations lacking a symmetry axis. Arrows indicate the directions of increasing $\gamma$ and $\eta$, which are generally opposite, except for the radial image in the cusp configuration (bottom-right inset of panel d).}
    \label{fig:geometry-cosmo}
\end{figure}

The cosmological constraining power of strong lensing depends not only on the number of sources across redshift but also on correlations between the lens mass profile and cosmology. In particular, the slope-cosmology degeneracy can be the dominant limiting factor in some systems. Here we show, however, that this degeneracy is strongly dependent on the image configuration.

For instance, in a double Einstein ring configuration, each ring forms where the convergence satisfies $\langle \kappa \rangle (<\theta) = 1$. Increasing the slope of the mass profile---while keeping the inner Einstein ring fixed---causes the outer ring to move inward, as $\kappa(\theta)$ decreases more steeply with radius. A similar effect occurs when lowering the matter density $\Omega_m$: the corresponding deflection ratio $\eta$ decreases, which also brings the two rings closer together. Thus, variations in slope and cosmology have qualitatively similar effects on the relative Einstein radii, making them difficult to disentangle from such configurations alone.

However, specific image configurations can help break this degeneracy, as shown in Fig.~\ref{fig:geometry-cosmo}. The figure compares the effect of varying $w$CDM parameters---which varies the deflection ratio $\eta(\Omega_m, w)$---with that of changing the EPL slope $\gamma$.

As shown in the first panel \ref{fig:geometry-cosmo-quad}, for a quadruply imaged system (and, by extension, for an Einstein ring), the effects of varying $\Omega_m$ and $w$ are nearly indistinguishable from those of changing $\gamma$, apart from minor differences in image shape. In such configurations, these parameters are strongly correlated, and the cosmological constraints are therefore limited by the mass profile slope.

In contrast, for a double image system (second panel \ref{fig:geometry-cosmo-double}), the impact of slope and cosmology changes differs significantly. The inner image, located closer to the core and radially (or nearly isotropically) magnified, is largely insensitive to the slope, while the outer, tangentially extended image is more affected. Both images, however, respond similarly to cosmology. This asymmetry in sensitivity helps to reduce the slope-cosmology degeneracy.

In cusp or fold configurations (panels \ref{fig:geometry-cosmo-cusp}, \ref{fig:geometry-cosmo-fold-slope}, and \ref{fig:geometry-cosmo-fold-cosmo}), the sensitivity differences are subtler: some images respond more strongly to changes in slope than others, which helps to mitigate the degeneracy. 

An especially interesting case arises when a cusp configuration produces a radial arc. In this case, variations in slope cause asymmetric shifts: the radial image moves outward (inward) as $\gamma$ increases (decreases), while the tangential images shift in the opposite direction. By contrast, changes in cosmology move all images inward or outward together. A similar effect occurs when the source of a double-image system lies near a cusp, again introducing asymmetries in the response to slope changes. Radial images are known to provide strong constraints on the inner slope of cluster-scale lenses \citep{Jullo2007, Robertson2019, Vega-Ferrero2021}, and this asymmetric behavior enables simultaneous constraints on both slope and cosmology with reduced correlation.

Radial images are rare in galaxy-scale lenses due to their steep inner mass profiles, but they occur more frequently in group- and cluster-scale lenses. Combined with the larger number of source planes in these systems, radial images make such lenses particularly valuable for multi-plane strong-lensing studies aimed at constraining dark matter and dark energy.

The Carousel Lens exhibits several of the strong-lensing configurations discussed above. Source~4 forms a quadruply imaged system that serves as a reference and primarily constrains the Einstein radius, with a fifth radial image (Paper~I). Source~1 forms a double-image system, with the source lying close to a cusp of the caustic. Source~7 produces a radial arc. In addition, sources~4 and~5 lie at the same redshift but are imaged at different positions in the lens plane. Together, these systems illustrate the range of image configurations present in the Carousel Lens that reduces the degeneracy between the mass slope and the cosmology, and motivate its suitability for multi–source-plane strong-lensing analysis. Finally, in a follow-up cosmological analysis, we will incorporate sources 12 and 13, both of which have radial images.

In the domain of galaxy-scale double-source-plane lenses, LSST and Euclid are expected to discover approximately 1700 such systems, respectively \citep{Sharma2023}. Early Euclid results have already identified four high-confidence candidates within $63 \, \text{deg}^2$, along with a larger number of lower-confidence candidates \citep{EuclidDSP2025}. Prioritizing follow-up observations for systems in which the second source produces double, cusp, or fold configurations may yield stronger constraints on cosmological parameters than focusing on double Einstein rings or double quads.


\newpage
\section{Forecast for Cosmological Constraints with $z > 4$ Sources}
\label{apx:additional-sources-forecast}
\begin{figure}[htb]
    \centering
    \includegraphics[width=0.35\linewidth]{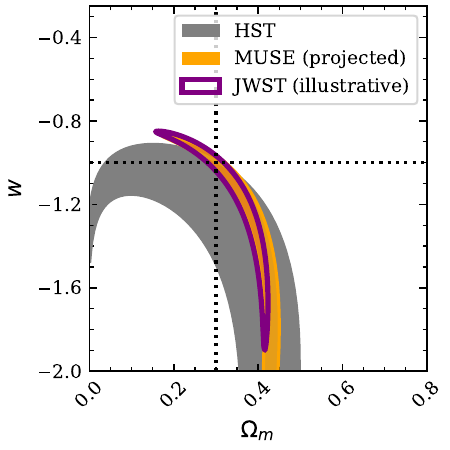}
    \caption{Projected $w$CDM constraints from the Carousel Lens with additional JWST high-redshift sources. 68\% credible contours from the current HST sources 1–7 (gray), projected constraints including the MUSE sources 8, 11, and 12 (orange), and an illustrative forecast incorporating hypothetical JWST sources at $z = 5$ and $z = 7$ (purple).
    }
    \label{fig:wCDM-JWST}
\end{figure}

While this work is primarily an analysis of the currently available data, rather than a full forecast study, we nevertheless include a brief illustration of the cosmological leverage that could be achieved with modest extensions of the present system. Even with the currently modeled set of sources, the Carousel Lens already provides competitive constraints on cosmological parameters, but its full potential is likely to be realized with additional sources. The depth and angular resolution of facilities such as JWST, and in the near future Roman, the ELTs \citep{skidmore2015, davies2016, males2022}, and concepts like Lazuli \citep{roy2026}, will discover new sources at $1.7<z<3$ and $z\gtrsim 4$ in the central lensing region with near certainty. 
As this work has demonstrated, this region is particularly powerful because of the existing 10 sources. 
Preliminary forecasts indicate that even the addition of just two higher-redshift sources, e.g., at $z=5$ and $z=7$, would substantially tighten the credible regions in the $w$–$\Omega_{\rm m}$ plane (Fig.~\ref{fig:wCDM-JWST}), under the assumption that our current model represents the true lens potential (which we will test against alternative model assumptions) and that multi-plane effects remain subdominant (which we intend to incorporate into future modeling). We emphasize that this section is intended only to illustrate the approximate ceiling of cosmological precision achievable with a relaxed, well-characterized multi-source-plane cluster strong-lensing system. A dedicated forecast study, incorporating realistic source populations and observational strategies, is in preparation.

\end{document}